\documentclass[review]{elsarticle}

\usepackage{lineno,hyperref}
\usepackage{framed,multirow}

\usepackage{amssymb}
\usepackage{latexsym}

\usepackage{url}
\usepackage{xcolor}
\usepackage[ruled, lined, linesnumbered, longend]{algorithm2e}
\usepackage{subfigure}
\usepackage{textcomp}
\usepackage{verbatim}

\modulolinenumbers[1]

\journal{Journal of \LaTeX\ Templates}

\begin{document}

\begin{frontmatter}

\title{Direct numerical simulations of incompressible multiphase electrohydrodynamic flow with single-phase transportation schemes}

\author[mymainaddress]{Qiang Liu}
\ead{qiangliu.7@outlook.com}
\author[mysecondaryaddress]{Jie Zhang}
\ead{j_zhang@xjtu.edu.cn}
\author[mymainaddress]{Jian Wu\corref{mycorrespondingauthor}}
\cortext[mycorrespondingauthor]{Corresponding author}
\ead{jian.wu@hit.edu.cn}

\address[mymainaddress]{School of Energy Science and Engineering, Harbin Institute of Technology, Harbin 150001, PR China}
\address[mysecondaryaddress]{School of Aerospace Engineering, Xi’an Jiaotong University, Xi’an 710049,PR China}

\begin{abstract}
In the present study, two schemes named face discernment and flux correction are proposed to achieve single-phase transportation of free charge in multiphase electrohydrodynamic(EHD) problems. Many EHD phenomena occur between air and another liquid while the free charge can only be transported in the liquid phase through ohmic conduction and convection due to the poor conductivity of air. However, the charge may be leaked into the dielectric air during the simulation due to the asynchronous transportation between interface and free charge.
To avoid this unphysical error, a face discernment method is designed to produce an accurate ohmic conduction of free charge by providing a superior physical properties distribution at the interface. Subsequently the flux correction method is developed to correct the advection flux of charge density to prevent ions crossing the interface. These two schemes are based on the Volume of Fluid (VOF) model and independent with the specific interface updating method. The performance of the proposed methods are carefully validated with several test cases. The algorithms are implemented as an OpenFOAM extension and are published as open source.
\end{abstract}


\end{frontmatter}

\section{\label{sec:introduction}Introduction}
Multi-phase electrohydrodynamics (EHD) portrays the interface flow under the electric field. When an external electric field is applied, the molecules in fluid will first be polarized, generating dipoles and inducing the dielectric force at the interface. The dielectric force only works in the normal direction of the interface if the fluid is isotropic and the polarization is homogeneous\cite{castellanos1998electrohydrodynamics,2006Electrodynamics}. Furthermore, the free charge inside the fluid will also be transported under the effect of ohmic conduction for conducting fluids. The migrated free charge will be accumulated at the interface due to the conductivity difference between the two fluids. The Coulomb force acting on these surface charges brings along an additional tangential component along the interface\cite{melcher1969electrohydrodynamics,Saville2003ELECTROHYDRODYNAMICS}. The Coulomb force and the dielectric force work as the main driving factor for the fluid flow under the electric field. In turn, the fluid flow also changes the distribution of dielectric and Coulomb forces by the interface deformation and convective charge transportation.

The combination of electric field and two-phase flows has always been a prosperous topic in the scientific community. On the one hand, it attracts a wide range of fundamental research interest due to its complex flow patterns and rich bifurcations\cite{Diversity2021,Vlahovska2019,Papageorgiou2019}. On the other hand, many industrial applications like electrosprays\cite{Ganan-Calvo2018,ROSELLLLOMPART20182}, EHD pumping\cite{HOJJATI2011294,2016Electrohydrodynamic} and fluid control in microfluidic devices\cite{YU2019114178,AZIZIAN2019201} all involve multiphase EHD flow. Although experiments\cite{Taylar1965,ha_yang_2000,SHIN200109955} and theoretical analysis\cite{melcher1969electrohydrodynamics,Saville2003ELECTROHYDRODYNAMICS,schnitzer2015taylor} have provided many paramount understandings to this area, numerical simulation still plays a vital role since it can provide information of all physical fields and manifest the details of the flow behavior.

Massive efforts have been made to develop the numerical tool for multiphase EHD problems. The earliest attempts started with the boundary element method\cite{sherwood_1988,baygents_rivette_stone_1998,SETIAWAN1997243,Higuera2007Emission}, which can only deal with the stokes flow and inviscid flow. Later works successfully broke through the restriction of flow type by introducing finite element method\cite{collins_harris_basaran_2007} and Lattice Boltzmann method\cite{ZHANG2005150} to solve the flow field. The electric force working at the interface and the interface charge transportation also bring challenges for the numerical modeling of the two-phase flow. A lot of efforts have been made to use different interface tracking models like the level-set method\cite{Sharp2009}, CLSVOF method\cite{Welch2007}, front-tracking method\cite{effects2015}and ghost fluid method\cite{earlyNumerical5} to describe the electric field coupled interface motion. Some of these algorithms become relatively complex when dealing with multiphase EHD problems and hard to implement in other simulation platforms, which limits their applications. Besides, most of the studies only focus on the EHD flow involving specific fluids like perfectly conducting fluid, perfectly dielectric fluid or leaky dielectric fluid\cite{melcher1969electrohydrodynamics,Saville2003ELECTROHYDRODYNAMICS}. A breakthrough was finally made by L\'opez-Herrera et.al\cite{Herrera} who proposed a finite volume method(FVM) based method and use the Volume of Fluid(VOF) method to track the interface. L\'opez-Herrera's method is capable of handling fluids with various conductivity and quite easy to implement. Many studies have been conducted based on this method\cite{Herrada2012,Lima2014,Roghair2015,HUANG20191,Dastourani2021}.

However, we have also noticed some flaws in recent practices of L\'opez-Herrera's method. For example, a charge density leakage has been obtained by Wirz et.al\cite{Wirz2019} during the simulation of the Taylor cone-jet process. The charge at the interface was transported to the insulating air and caused a non physical velocity in air under the action of Coulomb force. The root of this leakage is that the interface and the charge density are not transported synchronously in the numerical procedure and the relative motion between them leads to artificial transportation of charge density. This error can be ignored in some cases, but it will lead to non-physical leakage of charge when one phase is insulating. Most two-phase EHD phenomena with air belong to this category since the conductivity of air is often much lower than the liquid, and thus this error should be eliminated. Note that other problems involving two-phase scalar transport may also face similar challenges when the interface is tracked by solving a transportation equation like the procedure in VOF method. Examples include the heat transfer between two phases\cite{HILL2018285,MALAN2021109920} and conjugate mass transfer in chemical engineering\cite{Haroun2010,YANG20052643}. The conventional solution is to use the same advection flux for both the interface transportation and scalar transportation\cite{Lopez-Herrera2015,BOTHE2013283}. However, the charge is migrated by both convection and ohmic conduction while the interface is only advected by fluid flow, which brings along difficulty for the numerical coincident transportation procedure of charge and interface. Besides, numerous interface transport methods have been developed to ensure the sharpness and boundness of interface transport\cite{KATOPODES2019766,MOHAN2021110663,COMMINAL2021110479}, and adapting the charge transport scheme for each interface update method is unwise. Thus, it is necessary to develop a universal method to keep the free charge away from insulating phase.

The present work proposes two methods named as the face discernment and flux correction method to prevent the charge from leaking into the insulating phase. The face discernment method is inspired by the sharp scheme used in the previous magnetohydrodynamics studies\cite{Zhang2018} to confine the charge in the conducting phase, and it can give an accurate physical properties distribution near the interface. The charge accumulated at the interface can be guaranteed to only exist at the interface cells under the effect of face discernment method. The flux correction method is designed to ensure that the interface charge moves synchronously with the interface by correcting the velocity flux in charge transportation process. Both the face discernment method and the flux correction method are independent with the interface advection scheme, and thus they can be easily transplanted into various numerical platforms. In this study, the whole algorithm is implemented as an extension of the open source FVM framework OpenFOAM\cite{Weller1998} and the related code of this research is also released as open source.

The remainder of the paper is organized as follows: the description of the governing equations is presented in the next section; the numerical scheme and the implementation details of the proposed method are described in Section~\ref{sec::numericalMethod}; several cases are presented to test the present numerical method in Section~\ref{sec::ResultandDiscussion}; finally, the concluding remarks are summarized in the last section.

\section{Governing Equations}

For the present multiphase EHD flows, both the liquid and air are considered to be incompressible, Newtonian fluid. The governing equations for the flow motion consists of the continuity and momentum equations:
\begin{equation}
	\nabla \cdot \mathbf{u}
	=
	0
	\
	,
	\label{eq::continuity}
\end{equation}
\begin{equation}
	\frac{\partial \rho \mathbf{u}}{\partial t}
	+
	\nabla \cdot(\rho \mathbf{u u})
	=
	-\nabla p
	+
	\rho \mathbf{g}
	+
	\mathbf{F}_{\sigma}
	+
	\nabla \cdot \mathbb{T}_{\mu}
	+
	\nabla \cdot \mathbb{T}_{e}
	\label{eq::momentum}
\end{equation}
where $\mathbf{u}$ is the fluid velocity, $\rho$ is the fluid density and $p$ is the pressure. The term $\rho\mathbf{g}$ with gravitational acceleration $\mathbf{g}$ and $\mathbf{F}_{\sigma}$ refer to the gravity and surface tension, respectively. Besides, $\mathbb{T}_{\mu}$ in Eq.~(\ref{eq::momentum}) is the viscous stress and $\mathbb{T}_{e}$ is the Maxwell stress of electric field:
\begin{equation}
	\mathbb{T}_{\mu}
	=
	\mu\left(\nabla \mathbf{u}
	+
	\nabla \mathbf{u}^{T}\right)
	\
	,
	\label{eq::viscosStress}
\end{equation}
\begin{equation}
	\mathbb{T}_{e}
	=
	\varepsilon
	\left(
	\mathbf{E} \mathbf{E}
	-
	\frac{E^{2}}{2} \mathbb{I}
	\right)
	\label{eq::maxwellStress}
\end{equation}
where $\mu$ is dynamic viscosity, $\varepsilon$ is permittivity, $\mathbb{I}$ is the unit tensor and $\mathbf{E}$ is the electric field strength which can be obtained by
\begin{equation}
	\mathbf{E}
	=
	-\nabla \phi
	\
	.
	\label{eq::electricField}
\end{equation}
Here, $\phi$ is the electric potential derived from Poisson equation:
\begin{equation}
	\nabla \cdot
	\left(
	\varepsilon \nabla \phi
	\right)
	=
	-\rho_{e}
	\label{eq::Poisson}
\end{equation}
where $\rho_e$ represents the charge density. Substituting Eq.~(\ref{eq::electricField}) into Eq.~(\ref{eq::Poisson}) leads to another form of Eq.~(\ref{eq::Poisson}):
\begin{equation}
	\nabla \cdot \mathbf{D}
	=
	\nabla \cdot\left(
	\varepsilon \mathbf{E}
	\right)
	=
	\rho_{e}
	\label{eq::displacement}
\end{equation}
where $\mathbf{D}=\varepsilon\mathbf{E}$ is the electric displacement vector. The charge density in Eq.~(\ref{eq::Poisson}) and Eq.~(\ref{eq::displacement}) satisfies the following conservation equation\cite{Herrera}:
\begin{equation}
	\frac{\partial \rho_{e}}{\partial t}
	+
	\nabla \cdot\left(
	K \mathbf{E}
	\right)
	+
	\nabla \cdot\left(
	\rho_{e} \mathbf{u}
	\right)
	=
	0
	\label{eq::chargeConservtion}
\end{equation}
where $K$ is the conductivity. In Eq.~(\ref{eq::chargeConservtion}), $K\mathbf{E}$ and $\rho_{e} \mathbf{u}$ represent the ohmic conduction and flow convection component of charge transportation, respectively. Meanwhile, the divergence of $\mathbb{T}_{e}$ in Eq.~(\ref{eq::momentum}) can be treated as a volumetric electric force $\mathbf{F_e}$:
\begin{equation}
	\mathbf{F}_{\mathbf{e}}
	=
	\nabla \cdot \mathbb{T}_{e}
	=
	\nabla \cdot\left(
	\varepsilon \mathbf{E} \mathbf{E}
	-
	\frac{\varepsilon E^{2}}{2} \mathbb{I}
	\right)
	=
	\rho_{e} \mathbf{E}
	-
	\frac{1}{2} E^{2} \nabla \varepsilon
	\label{eq::electricForce}
\end{equation}
where $\rho_{e} \mathbf{E}$ is the Coulomb force and $E^{2} \nabla \varepsilon/2$ represents the dielectric force.

Due to the imposed electric field, the stress balance at the interface can be described as\cite{Herrera,Chicon2006}:
\begin{equation}
	\mathbf{t} \cdot\left[\mathbb{T}_{\mu}\right] \cdot \mathbf{n}
	+
	\mathbf{t} \cdot\left[\mathbb{T}_{e}\right] \cdot \mathbf{n}
	=
	0
	\
	,
	\label{eq::interStressT}
\end{equation}
\begin{equation}
	{[p]+\mathbf{n} \cdot\left[\mathbb{T}_{\mu}\right] \cdot \mathbf{n}
		+
		\mathbf{n} \cdot\left[\mathbb{T}_{e}\right] \cdot \mathbf{n}
		=
		\sigma \nabla_{s} \cdot \mathbf{n}}
	\label{eq::interStressN}
\end{equation}
where
\begin{equation}
	\mathbf{t} \cdot\left[\mathbb{T}_{e}\right] \cdot \mathbf{n}
	=
	\rho_{e, s}(\mathbf{E} \cdot \mathbf{t})
	\
	,
	\label{eq::maxwellT}
\end{equation}
\begin{equation}
	\mathbf{n} \cdot\left[\mathbb{T}_{e}\right] \cdot \mathbf{n}
	=
	\left[
	\varepsilon(\mathbf{E} \cdot \mathbf{n})^{2}
	\right]
	-
	\left[
	\frac{1}{2} \varepsilon E^{2}
	\right]
	\
	.
	\label{eq::maxwellN}
\end{equation}
Here, $\sigma$ is the surface tension coefficient, $\left[ A\right] =A_1-A_2$ represents the jump of any quantity $A$ from phase 1 to phase 2, $ \nabla_{s}$ is the surface gradient operator\cite{melcher1969electrohydrodynamics,Castellanos1998} , $\rho_{e, s}$ is the interface charge density, $\mathbf{n}$ and $\mathbf{t}$ are the unit normal and tangential vector of the interface, respectively. Since the dielectric force only acts in the normal direction of the interface, it appears in Eq.~(\ref{eq::maxwellN}) while vanishes in Eq.~(\ref{eq::maxwellT}). The derivation in Eq.~(\ref{eq::maxwellT}) and Eq.~(\ref{eq::maxwellN}) also involves the following relationship derived by applying Eq.~(\ref{eq::displacement}) at the interface:
\begin{equation}
	\left[\mathbf{D}\right] \cdot \mathbf{n}
	=
	\left[\varepsilon \mathbf{E}\right]\cdot \mathbf{n}
	=
	\rho_{e, s}
	\
	.
	\label{eq::electricFieldInter}
\end{equation}
The $\rho_{e, s}$ here satisfies the following conservation equation\cite{Herrera}:
\begin{equation}
	\frac{\partial \rho_{e, s}}{\partial t}
	+
	\mathbf{u} \cdot \nabla_{s} \rho_{e, s}
	+
	\left[
	K \mathbf{E} \cdot \mathbf{n}
	\right]
	=
	\rho_{e, s} \mathbf{n} \cdot\left(
	\mathbf{n} \cdot \nabla
	\right) \cdot \mathbf{u}
	\label{eq::interfaceCharge}
\end{equation}
where the R.H.S represents the charge transportation due to the movement of the interface.

To track the interface, the VOF model\cite{VOF} is introduced. A scalar transport equation for the phase fraction $\alpha$ can be given as:

\begin{equation}
	\frac{\partial \alpha}{\partial t}
	+
	\nabla \cdot\left(
	\alpha \mathbf{u}
	\right)
	=
	0
	\label{eq::VOF}
\end{equation}
where the value of $\alpha$ varies from 0 to 1, and the regions with $\alpha=1$ and $\alpha=0$ are marked as phase 1 and phase 2, respectively. Consequently, the physical properties are expressed as a function $f_a$ of local phase fraction:
\begin{equation}
	P
	=
	f_{a}\left(
	\alpha, P_{1}, P_{2}
	\right)
	\label{eq::propertiesAverage}
\end{equation}
where ``$P$'' represents physical properties including $\rho$, $\mu$, $K$, and $\varepsilon$, the subscripts ``1'' and ``2'' indicate the values of the phase 1 and phase 2, respectively. The selection of $f_a$ is a controversial topic in multiphase EHD research\cite{Herrera,Tomar2007}. In this study, two different average methods named the linear average method and the harmonic average method are also involved to validate their performance in the simulation:
\begin{equation}
	f_{a,{ linear }}\left(
	\alpha, P_{1}, P_{2}
	\right)
	=
	\alpha P_{1}
	+
	(1-\alpha) P_{2}
	\
	,
	\label{eq::linearAverage}
\end{equation}
\begin{equation}
	f_{a,{ average }}\left(
	\alpha, P_{1}, P_{2}
	\right)
	=
	\frac{P_{1} P_{2}}{\alpha P_{2}
		+\
		\left(
		1-\alpha
		\right) P_{1}}
	\
	.
	\label{eq::harmonicAverage}
\end{equation}

\section{Numerical Method\label{sec::numericalMethod}}
\subsection{\label{sec::siscretisation}Discretisation of electric equations}
The proposed algorithm is built as an extension of the open source FVM framework of OpenFOAM. The discretisation procedure of the Navier-stokes equations Eq.~(\ref{eq::continuity}) and (\ref{eq::momentum}) in OpenFOAM has already been well discussed in many literatures\cite{Weller1998,Jasak,moukalled2016finite}, and thus only the discretisation of the electric equations are explained in this section.

With the standard FVM discretisation procedure, the Possion equation Eq.~\ref{eq::Poisson} and charge conservation equation Eq.~(\ref{eq::chargeConservtion}) are firstly discretized using the following time marching scheme respectively:
\begin{equation}
	\frac{1}{V_{p}} \sum_{f}\left[
	\varepsilon_{f}^{n}\left(
	\nabla \phi^{n}
	\right)_{f} \cdot \mathbf{S}_{\mathbf{f}}
	\right]
	=
	-\rho_{e, c}^{n-1}
	\
	,
	\label{eq::disPossion}
\end{equation}
\begin{equation}
	\frac{3 \rho_{e, c}^{n}-4 \rho_{e, c}^{n-1}+\rho_{e, c}^{n-2}}{2 \Delta t} V_{p}
	+
	\sum_{f}\left(
	\rho_{e, f}^{n} \mathbf{u}_{f}^{n-1} \cdot \mathbf{S}_{\mathbf{f}}
	\right)
	=
	\sum_{f}\left[
	K_{f}^{n}\left(\nabla \phi^{n}\right)_{f} \cdot \mathbf{S}_{\mathbf{f}}
	\right]
	\label{eq::disChargeConservation}
\end{equation}
where $\Delta t$ is the time step, $V_p$ refers to the volume of the mesh cell and $\mathbf{S_f}$ is the surface vector.
All the values are stored at the center of the cell with the subscript ``$c$'' by default except those face-centered variables with the subscript ``$f$''. Besides, the superscripts ``$n$'',``$n-1$'' and ``$n-2$'' denote the variables at the present and two previous time steps. The discretisation scheme of the transient term in Eq.~(\ref{eq::disChargeConservation}) is also known as three-time-level backward scheme\cite{backward}. To keep concise, the superscript will not be specifically marked in the following content if the variables adopt the same discrete scheme at all time steps.

In this study, the electric field strength $\mathbf{E}$ is not directly calculated from electric potential $\phi$ by the standard finite-volume procedure as $\mathbf{E_c}=-\nabla\phi_c=-\sum_f\phi_f\mathbf{S_f}/V_p$, but from the electric displacement $\mathbf{D}$ using a reconstruction scheme proposed by R. Thirumalaisamy et al\cite{Thirumalaisamy2018}:
\begin{equation}
	\mathbf{D}_{c}
	=
	-\frac{1}{V_{p}} \sum_{f} \varepsilon_{f}(\nabla \phi)_{f}\left(\mathbf{C}_{f}-\mathbf{C}_{c}\right) \cdot \mathbf{S}_{\mathbf{f}}
	\
	,
	\label{eq::disDisplacement}
\end{equation}
\begin{equation}
	\mathbf{E}_{\mathbf{c}}
	=
	\frac{\mathbf{D}_{\mathbf{c}}}{\varepsilon_{c}}
	\label{eq::disElectricField}
\end{equation}
where $\mathbf{C}_{f}$ and $\mathbf{C}_{c}$ are the position vector of the face center and cell center of a control volume, respectively. Then, the divergence of Maxwell stress can be calculated as\cite{Thirumalaisamy2018}
\begin{equation}
	\mathbf{F}_{\mathbf{e}, \mathbf{c}}
	=
	\nabla \cdot\left(
	\varepsilon_{c} \mathbf{E}_{\mathbf{c}} \mathbf{E}_{\mathbf{c}}-\frac{\varepsilon_{c} E_{c}^{2}}{2} \mathbb{I}
	\right)
	=
	\frac{1}{V_{p}} \sum_{f}\left[
	\mathbf{D}_{\mathbf{f}}\left(\mathbf{E}_{\mathbf{f}} \cdot \mathbf{S}_{\mathbf{f}}\right)-\frac{\left(D^{2}\right)_{f}}{2 \varepsilon_{f}} \mathbf{S}_{\mathbf{f}}
	\right]
	\label{eq::disFeOld}
	\
	.
\end{equation}
This reconstruction scheme is proved to give a more accurate electric force distribution near the interface\cite{Thirumalaisamy2018}. In our method, a slight modification is made to Eq.~(\ref{eq::disFeOld}):
\begin{equation}
	\frac{1}{V_{p}} \sum_{f}\left[
	\mathbf{D}_{\mathbf{f}}\left(\mathbf{E}_{\mathbf{f}} \cdot \mathbf{S}_{\mathbf{f}}\right)
	-
	\frac{\left(D^{2}\right)_{f}}{2 \varepsilon_{f}} \mathbf{S}_{\mathbf{f}}
	\right]
	\Rightarrow
	\frac{1}{V_{p}} \sum_{f}\left[
	\mathbf{D}_{\mathbf{f}}\left(\frac{\mathbf{D}_{\mathbf{f}} \cdot \mathbf{S}_{\mathbf{f}}}{\varepsilon_{f}}\right)
	-
	\frac{\left(D^{2}\right)_{f}}{2 \varepsilon_{f}} \mathbf{S}_{\mathbf{f}}
	\right]
	\
	.
	\label{eq::disFeNew}
\end{equation}
Here, the $\mathbf{E_f}\cdot\mathbf{S_f}$ is replaced by $\mathbf{D_f}\cdot\mathbf{S_f}/\varepsilon_{f}$ in Eq.~(\ref{eq::disFeNew}). The reason for this modification is that $\mathbf{E_f}$ involves an face interpolation of volumetric $\varepsilon_c$ since the electric field is obtained by displacement as shown in Eq.~(\ref{eq::disElectricField}). As will be discussed in the coming section, the face interpolation of the physical properties is inaccurate near the interface. While the proposed face discernment method is capable to provide a more accurate surface value of permittivity, and thus the $\varepsilon_f$ is introduced into Eq.~$\ref{eq::disFeNew}$.

The full list of the interpolation schemes used to obtain the face values in Eq.~(\ref{eq::disPossion})-(\ref{eq::disFeNew}) are showed in Table \ref{tab::schemesTable}.

\begin{table}[htbp]
	\centering
	\caption{The interpolation schemes used in the discretisation procedure}
	\label{tab::schemesTable}
	\begin{tabular*}{1\textwidth}{@{\extracolsep{\fill}}ccc}
		\hline
		& Term & Schemes \\
		\hline
		\multirow{5}{*}{Electric equations}& $\varepsilon_f$ & face discernment \\
		
		& $K_f$ & face discernment \\
		
		& $\left(\nabla\phi\right)_f\cdot\mathbf{S_f}$ & corrected\cite{Jasak,correctedScheme} \\
		
		& $u_f$ & linear \\
		
		& $\rho_{e,f}$ & vanLeer\cite{van1997towards} \\
		
		&  &  \\
		
		\multirow{4}{*}{Other Equations$^+$}& transient term & backward\cite{backward} \\
		
		& convection term & QUICK\cite{leonard1979stable} \\
		
		& difussion term$^*$ & linear corrected \\
		
		& gradient term & pointCellsLeastSquares\cite{LeastSquaresGrad} \\
		\hline
		\multicolumn{3}{@{}l}{{\footnotesize $^+$ Other equations include Navier-stokes equations and the equations related to the interface }}\\
		\multicolumn{3}{@{}l}{{\footnotesize update. }}\\
		\multicolumn{3}{@{}l}{{\footnotesize $^*$ The finite volume discretization of the diffusion term for any quantity $Q$ with the diffusion}}\\
		\multicolumn{3}{@{}l}{{\footnotesize coefficient D is $\nabla\left(D\nabla Q\right)=\left[\sum_f D_f\left(\nabla Q\right)\cdot\mathbf{S_f}\right]/V_p$ .Here, the linear scheme is used for $D_f$}}\\
		\multicolumn{3}{@{}l}{{\footnotesize and the corrected scheme is chossen for $\left(\nabla Q\right)\cdot\mathbf{S_f}$.}}
	\end{tabular*}
\end{table}
\subsection{Phase update and surface tension framework\label{sec::PhaseUpdate}}
The phase update procedure and surface tension calculation in the present method is based on the TwoPhaseFlow\cite{Scheufler2021} framework designed by Henning Scheuflera and Johan Roenby.

\begin{figure}[t]
	\centering
	\includegraphics[scale=0.7]{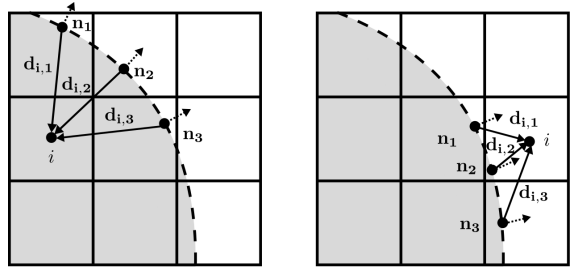}
	\caption{\label{fig::RDF}The sketch of the vectors used in the calculation of RDF with the dashed line indicating the interface. The shaded and blank areas represent the region filled with single phase.}
\end{figure}

The phase update method is chosen as the isoAdvector scheme\cite{Roenby2016} available in TwoPhaseFlow. In this scheme, the interface is firstly reconstructed inside a mesh cell based on the volumetric phase fraction. Then, the reconstructed interface is directly advected according to the local velocity interpolated to the interface position, and the volumetric phase fraction after the advection is calculated by integrating the submerged area of the new interface in a cell\cite{Roenby2016}. To reconstruct the interface for the isoAdvector method, a second order plicRDF scheme\cite{Scheufler2019} is introduced. This scheme extends the classical Piecewise
Linear Interface Construction(PLIC) scheme with the orientation computed by the gradient of the Reconstructed Distance Function(RDF) $\psi$\cite{Scheufler2019,CUMMINS2005425}:
\begin{equation}
	\psi_{i}
	=
	\frac{
		\sum_{j}\left(\mathbf{n}_{\mathbf{j}} \cdot \mathbf{d}_{\mathbf{i}, \mathbf{j}}\right)^{3}
		/
		\left|\mathbf{d}_{\mathbf{i}, \mathbf{j}}\right|^{2}
	}{
		\left(\mathbf{n}_{\mathbf{j}} \cdot \mathbf{d}_{\mathbf{i}, \mathbf{j}}\right)^{2}
		/
		\left|\mathbf{d}_{\mathbf{i}, \mathbf{j}}\right|^{2}
	}
	\
	.
	\label{eq::RDF}
\end{equation}
Here, the subscript $j$ represents the interface cell which share the same mesh nodes with the cell $i$, $\psi_i$ is the RDF in the centre of cell $i$ and $\mathbf{n_j}$ is the unit normal vector of the interface in cell $j$. The quantity $\mathbf{d_{i,j}}=\mathbf{C_{c,i}}-\mathbf{C_{interface,j}}$ is the distance vector between cell $i$ and the interface in cell $j$ where $\mathbf{C_{c,i}}$ and $\mathbf{C_{interface,j}}$ are the corresponding position vectors. Fig.~\ref{fig::RDF} illustrates the above variables in RDF calculation process.

To give a volumetric surface tension in momentum equation, the Continuous Surface Force(CSF)\cite{CSF} model is introduced as
\begin{equation}
	\mathbf{F}_{\sigma}
	=
	\sigma \kappa \nabla \alpha
	\label{eq::CSF}
\end{equation}
where $\kappa$ is the interface curvature which can be obtained by the divergence of unit interface normal vector $\mathbf{n}$. The gradient of RDF is also selected to validate the interface normal vector here, thus Eq.~(\ref{eq::CSF}) can be rewritten as\cite{Scheufler2021}:
\begin{equation}
	\mathbf{F}_{\sigma}
	=
	\sigma \kappa \nabla \alpha
	=
	\sigma\left(
	\nabla \cdot \mathbf{n}
	\right) \nabla \alpha
	=
	\sigma\left(
	\nabla \cdot \frac{\nabla \psi}{|\nabla \psi|}
	\right) \nabla \alpha
	\
	.
	\label{eq::surfaceTension}
\end{equation}

Although OpenFOAM provides a default method of calculating the interface normal vector using the gradient of phase fractions $\alpha$, the RDF model is proved to give a more accurate curvature estimation\cite{Roenby2016}.

\subsection{The face discernment method}

\begin{figure}[t]
	\centering
	\subfigure{
		\label{fig::phaseDistribution_a}
		\includegraphics[scale=0.7]{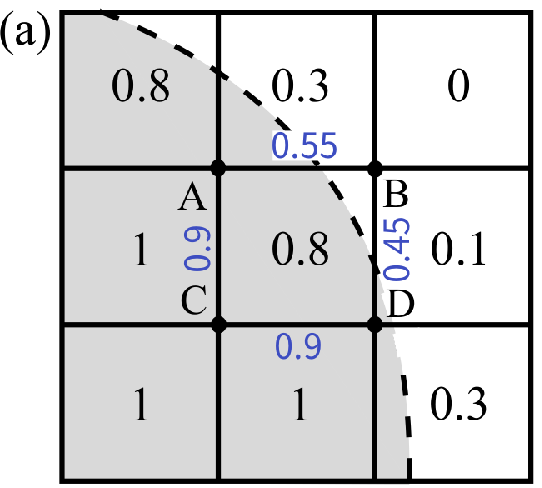}
	}
	\hspace{0.5em}
	\subfigure{
		\label{fig::phaseDistribution_b}
		\includegraphics[scale=0.7]{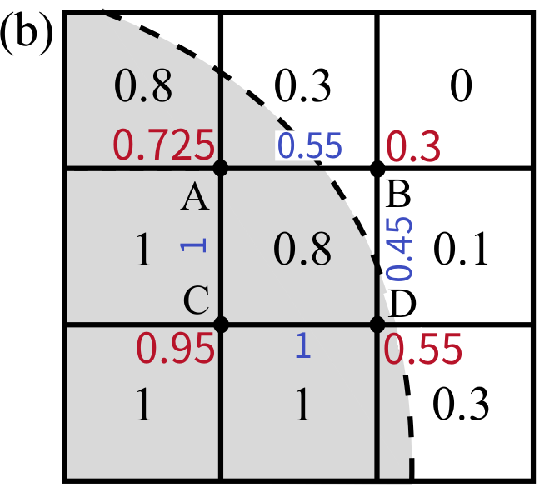}
	}
	\caption{\label{fig::phaseDistribution}The sketch of the phase fraction distribution on the one-phase faces and mixed-phase faces. The shaded and blank areas represent the region filled with single phase. The dashed line and the number in the cell represent the interface and local phase fraction, respectively.  A linear interpolation is used to compute face values. (a) The phase distribution without face discernment method. (b) The phase distribution with face discernment method.}
\end{figure}

In Section \ref{sec::siscretisation}, we show that the solution of the discretized equations, i.e. Eq.~{\ref{eq::disPossion}} and Eq.~{\ref{eq::disChargeConservation}}, strongly relies on the value of the permittivity ($\varepsilon_f$) and the conductivity ($K_f$) at the face center of the control volume. However, the linear- or harmonic-averaged algorithm for the estimation of $\varepsilon_f$ and $K_f$ at the face center, as described by Eq.~{\ref{eq::linearAverage}} and Eq.~{\ref{eq::harmonicAverage}}, may decline the conservative property of the charge densities. A typical two-dimensional example is illustrated in Fig.~\ref{fig::phaseDistribution}, where panel (a) shows how to compute $\varepsilon_f$ by using the original averaging scheme. In the sketch, the liquid and gas phases are separated by the isoline of $\alpha = 0.5$, as denoted by the dashed line. If we define $\varepsilon$ to be 1 in the liquid and 0 in the gas for simplicity, the embedded numbers in the cell center show the volume-averaged values of $\varepsilon$, then the face-centered $\varepsilon_f$ can be interpolated from the neighboring cells, denoted by the blue numbers in the same picture. Nevertheless, such interpolation may undermine the physical authenticity under some circumstances. For instance, we would expect $\varepsilon_f$ to be exactly 1 at the face of AC because it is fully immersed in the liquid phase, however, the interpolation scheme obtains 0.9 numerically. This deviation, undoubtedly, leads to the inaccurate estimations of the flux of charge densities at the cell face, and more seriously, spurious fluxes are produced at the pure gas faces which leads to the numerical leakage of charge densities, by which the conservativeness of the charge density is declined.

To fix this problem, we propose a method aiming at categorizing the cell faces into different types, denoted as ``face discernment scheme" in the following study. To begin with, we estimate the corner values of $\alpha_c$ by averaging them from the surrounding cells, denoted by the red numbers in Fig.~\ref{fig::phaseDistribution_b}, which is identical with that of panel (a). Then we classify the cell faces into two types: the one-phase faces and the mixed-phase faces, the former indicates the faces bounded by vertices both having $\alpha_c > 0.5$ or $\alpha_c < 0.5$, while the later implies the face is traversed by the isoline of $\alpha = 0.5$ so that the two vertices are respectively $\alpha_c > 0.5$ and $\alpha_c < 0.5$. According to the picture, clearly, AC and CD correspond to the one-phase face, while AB and BD belong to the mixed-phase face. Correspondingly, as shown in panel (b), we have assigned the deserved values of $\varepsilon_f$ after such identification. Keep in mind that the extension of this face discernment scheme can be extended to three-dimensional straightforwardly since there is no any indispensable two-dimensional precondition.

\begin{figure}[htbp]
	\centering
	\includegraphics[scale=0.7]{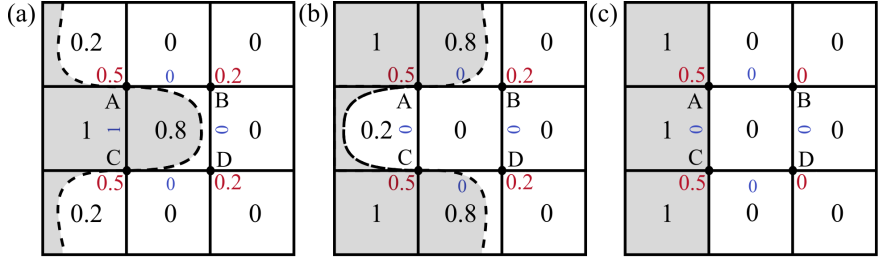}
	\caption{\label{fig::cutCell_a}\label{fig::cutCell_b} \label{fig::cutCell_c}\label{fig::cutCell}The sketch of the face phase fraction $\alpha_f$ obtained by face discernment methods in some extreme cases.  The shaded and blank areas represent the region filled with single phase. The dashed line and the number in the cell represent the interface and local phase fraction, respectively.  (a) The situation when the interface is convex to the region of $\alpha = 0$. (b) The situation when the interface is convex to the region of $\alpha = 1$. (c) The situation when the interface is the cell face.
	}
\end{figure}

Besides, there are some other extreme cases given that the phase fraction at the vertice is exactly $0.5$, it indicates the iso-line of $\alpha_c = 0.5$ crosses over the vertice precisely. Fig.~\ref{fig::cutCell} summarises the possible three topological structures of the interface on the discretised grids. To enhance the conservativeness of the charge densities, still, we must guarantee that the value of $\varepsilon_f$ at the face of AC is $1$ in panel (a), 0 in panel (b) and (c), respectively.

\subsection{The flux correction method}

The afore mentioned face discernment scheme enables us to correct the value of the permittivity and the conductivity at the cell faces, so that the flux of $\varepsilon_{f}\left(\nabla \phi\right)_{f} \cdot \mathbf{S}_{\mathbf{f}}$ and $K_{f}\left(\nabla \phi\right)_{f} \cdot \mathbf{S}_{\mathbf{f}}$ can be evaluated correctly by eliminating the possible spurious flux. Furthermore, the convective flux is also worthy of attention for the same reason, given as the discretisation of $\rho_{e,f}\mathbf{u}_f\cdot \mathbf{S}_{\mathbf{f}}$. A similar approach for correction is also designed by taking the phase fraction at the cell face into consideration. In brief, the convective flux is evaluated as $\alpha_f\rho_{e,f}\mathbf{u}_f\cdot \mathbf{S}_{\mathbf{f}}$ at the cell face, with $\alpha_f$ denoting the face-centered phase fraction after applying the face discernment scheme.  To simplify the expression, the flux $\alpha_f\rho_{e,f}\mathbf{u}_f\cdot \mathbf{S}_{\mathbf{f}}$ is named as single-phase correction flux. The single-phase correction flux guarantees that the charge will only by convected in the region where $\alpha\ne 0$ and thus further improves the numerical conservativeness of the charge densities in the conducting phase.

\begin{figure}[htbp]
	\centering
	\includegraphics[scale=0.7]{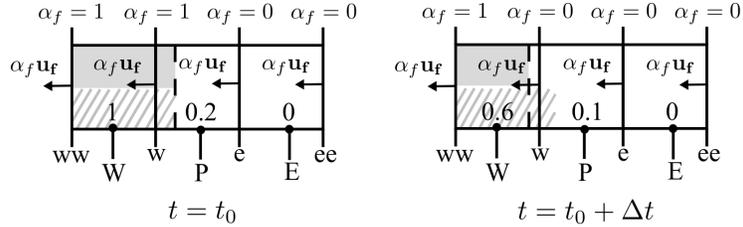}
	\caption{\label{fig::transportSketch}The sketch showing the stuck charge in the insulating phase. The shaded area and dashed areas represent the region filled with conducting phase($\alpha=1$) and the local charge density, respectively. The dashed line and the number in the cell represent the interface and detailed local charge density value, respectively.  
	}
\end{figure}

Nevertheless, another problem may arise by introducing the single-phase correction flux, as shown in Fig.~\ref{fig::transportSketch}. In the picture, the convection velocity $\mathbf{u}_f$ transports the charge density $\rho_e$ from right to left. At $t = t_0$, cell W is full of liquid as $\alpha_c = 1$, E corresponds to the full gas cell of $\alpha_c = 0$, and P is an interfacial cell with $\alpha_c = 0.2$ that the dashed line implies the interface position. Note that for the reason of simplicity, we just consider a one-dimensional problem here, so the interface is parallel to the cell faces, which are denoted by ww, w, e and ee in the pictures. Still at $t = t_0$, we assume $\rho_e^{t_0}$ has the value of 1, 0.2 and 0 in cells W, P and E respectively. The face centered $\rho_{e,f}$ is obtained by upwind scheme here for both simplicity and keeping the upwind feature of the vanLeer scheme shown in Tab.\ref{tab::schemesTable}. Considering the leftwards velocity, it implies the face centered charge density to be $\rho_{e,ww} = 1$ and $\rho_{e,w} = 0.2$ and $ \rho_{e,e} = 0$ when $t=t_0$. Then by assuming the velocity flux in one time step is a constant of $|\mathbf{u}_f\cdot \mathbf{S}_{\mathbf{f}}\Delta t|                                                                                                                                                                                                                                                                                                                                                                                                                         = 0.5$ at all the faces, the cell centered charge density at the next time step is calculated as $\rho_e^{t_0+\Delta t} = \rho_e^{t_0} + \sum\rho_{e,f}\alpha_f\mathbf{u}_f\cdot \mathbf{S}_{\mathbf{f}}\Delta t$, leading to $\rho_{e,W}^{t_0+\Delta t} = 0.6, \rho_{e,P}^{t_0+\Delta t} = 0.1, \rho_{e,E}^{t_0+\Delta t} = 0$. Since the interface has been moved to cell W at $t = t_0 + \Delta t$, the face centered phase fraction $\alpha_f$ at face w and e will equal to 0 according to the face discernment method, indicating that the charge density in cell P cannot be transported to cell W by the flux $\alpha_f\rho_{e, f}\mathbf{u}_f\cdot \mathbf{S}_{\mathbf{f}}$ in the following time steps and thus the $\rho_e$ in cell P is trapped. To overcome this problem, we define an additional correction flux of $c_f\rho_{e,f}\mathbf{u}_f\cdot \mathbf{S}_{\mathbf{f}}$ at the cell face, with $c_f$ denoting the correction coefficient to transport all the charge density in one cell to its neighboring cell if the interface is moving away. Hence the total convective flux becomes $\left( \alpha_f+c_f \right)\rho_{e,f}\mathbf{u}_f\cdot \mathbf{S}_{\mathbf{f}}$.

\begin{figure}[tb]	
	\centering
	\includegraphics[scale=0.7]{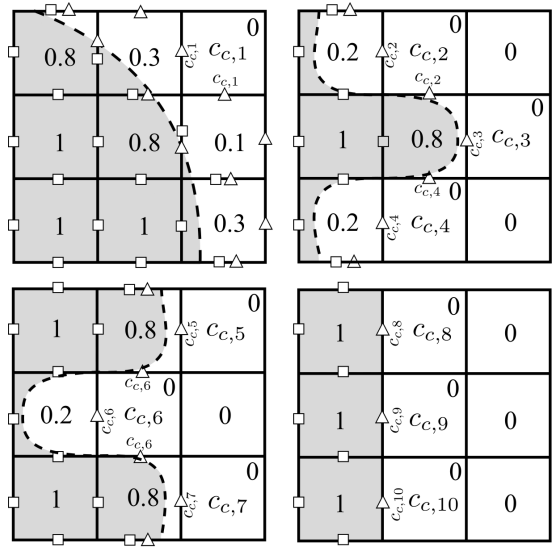}
	\caption{\label{fig::fluxCorrection}The distribution of single-phase correction flux and additional correction flux in the mesh.  The shaded and blank areas represent the region filled with single phase. The dashed line and the number in the cell represent the interface and local phase fraction, respectively.  The symbol ``$\Box$'' and ``$\triangle $'' at faces mark the faces where single-phase correction flux and additional correction flux work, respectively. 
	}
\end{figure}
The additional correction flux $c_f\rho_{e,f}\mathbf{u}_f\cdot \mathbf{S}_{\mathbf{f}}$ should only work at face w when $t=t_0+\Delta t$ in Fig.\ref{fig::transportSketch} to advect the charge in cell P. For a more general two- or three-dimensional case, the additional correction flux should locate at the faces of the interface cells whose $\alpha_f \ne1$, as shown in Fig.\ref{fig::fluxCorrection}. For the specific  value of $c_f$, it is natural to set the initial estimation as $c_f = 1$. However, this original value does not guarantee the charge density to be fully transported to the neighboring cell if the interface is moving away. For instance, the charge density variation in cell P (see Fig.~\ref{fig::transportSketch}) is $\Delta \rho_{e,P}^{t+\Delta t} = c_{f,w}\rho_{e,w}^{t+\Delta t}\mathbf{u}_w\cdot \mathbf{S}_{\mathbf{f}}\Delta t=-0.05$, which is not adequate to fully transport the charge density from cell P. Thus, a prediction-correction procedure is required to finalize the value of $c_f$ on the cell faces. In details, we firstly transport the charge density solely with $\alpha_f\rho_{e,f}\mathbf{u}_f\cdot \mathbf{S}_{\mathbf{f}}$, and we obtain the prediction value of $\rho_{e}^{t_0+\Delta t}$ in each cell. Then a prior estimation of $c_f=1$ is assigned, so the charge density variation is calculated as
\begin{equation}
	\Delta \rho_{e}
	=
	-\Delta t \frac{1}{V_{p}} \sum_{f}\left(c_f\rho_{e,f}\mathbf{u}_f\cdot \mathbf{S}_{\mathbf{f}}\right)
	\
	.
	\label{eq::deltaRhoe}
\end{equation}
Then to guarantee that $\rho_{e,P}^{t+\Delta t} = 0$, $c_f$ at cell face w in Fig.\ref{fig::transportSketch} could be corrected as $c_{f,w} = -\rho_{e,P}/\Delta \rho_{e,P}$. For the two- or three dimensional case, the initial estimation of $c_f$ is $c_f=1-\alpha_f$ and the correction coefficient at the faces where $\alpha_f=0$ is set to the value of the $c_{c,i}=-\rho_{e,i}/\Delta \rho_{e,i}$ where ``i''  is the index of the adjacent cell whose phase fraction is 0. Fig.~\ref{fig::fluxCorrection} depicts the detailed corresponding relationship between face centered $c_f$ and cell centered $c_{c,i}$ in different cases.

\subsection{Charge transportation in sub-grid droplets}
\begin{figure}[h]
\centering
\subfigure{
	\label{fig::subGrid_a}
	\includegraphics[scale=0.7]{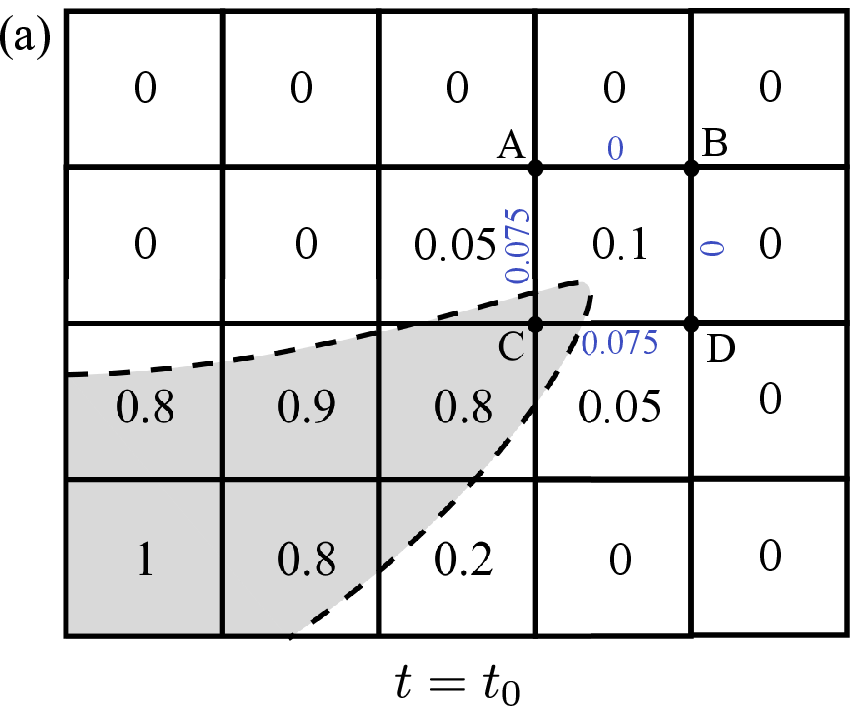}
}
\subfigure{
	\label{fig::subGrid_b}
	\includegraphics[scale=0.7]{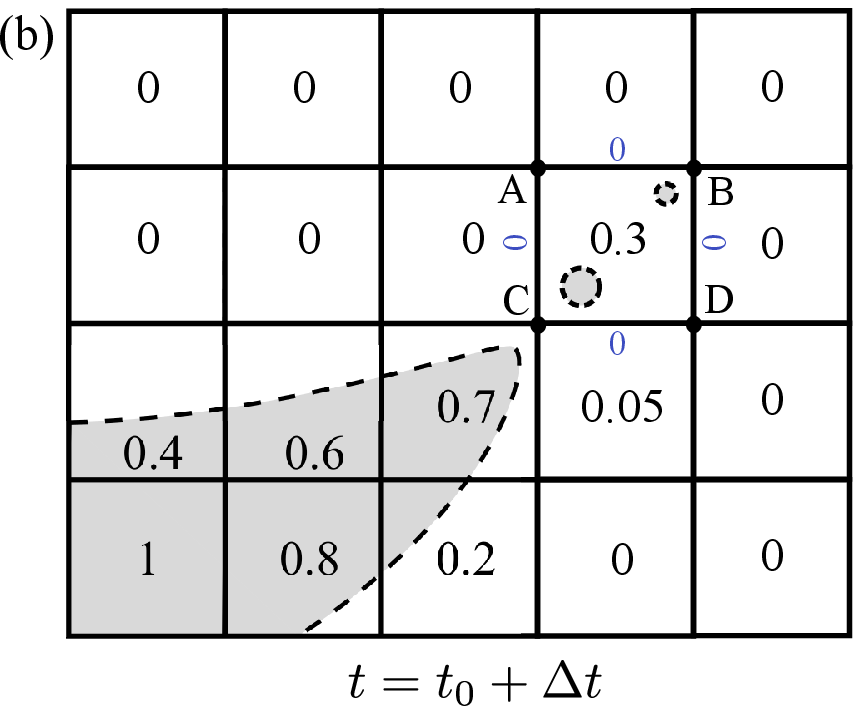}
}
\caption{\label{fig::subGrid}The sketch of a typical subgrid droplets generation process.  The shaded and blank areas represent the region filled with single phase. The dashed line and the number in the cell represent the interface and local phase fraction, respectively. 
}
\end{figure}

In this subsection, the procedure dealing with the charge in the sub-grid droplets will be discussed. As a result of the competition between the Maxwell stress, viscous stress and surface tension at the interface, some droplets with their size smaller than the mesh cell size may be formed during the simulation. A typical generation process of these sub-grid droplets are depicted in Fig.~\ref{fig::subGrid}. Since the interface doesn't cross the mesh faces at the cell where sub-grid droplets are located, the face-centred phase fraction $\alpha_f$ at those cell faces equal to 0, as the face A-C, C-D, D-B and B-A in Fig.~\ref{fig::subGrid_b}. This will result in a uniform distribution of conductivity and permittivity at the faces according to the face discernment method, which means the ohmic conduction of charge will vanish at the sub-grid droplets cell if the surrounding phase is insulated. Besides, since the $\alpha_f$ is 0 and the faces are not connected to interface cell, both the single-phase correction flux and additional correction flux will equal to 0 at those faces, thus the convection in charge transportation is also absent. As a result, the charge carried by these sub-grid droplets will be stuck in the cells due to the lack of conduction and convection. The Coulomb force applied on the stuck charge will destabilize the flow and thus this error need to be avoid.

Note that the charge in the sub-grid droplets will be successfully transported under the effect of ohmic conduction and convection if the face discernment method and flux correction method is not applied in the simulation procedure. Thus, the sub-grid charge is separated from the total charge at the beginning of each time step in the algorithm, and only the charge in the cells filled with conducting phase and the interface cells will be transported with face discernment and flux-correction method. For the sub-grid cells, Eq.~(\ref{eq::disPossion}) and Eq.~(\ref{eq::disChargeConservation}) are directly solved with the conductivity and permittivity interpolated from cell centres. Finally, the charge density will be reconstructed from the sub-grid charge and the charge in conducting phase and interface.

\subsection{Summary and the overall solution procedure}
In previous subsections, the details of the face discernment method and flux correction method have been presented. Although these two methods are implemented as an extension of OpenFOAM at present, the cost of transplanting them to other platforms are very low since the only input data required by these two method is a volumetric phase fraction distribution and nothing related to the specific interface updating method is introduced. One may also naturally doubt the conservation of the face discernment method and flux correction method since the physical properties and velocity flux is artificially modified. However, the proposed methods don't damage the conservation feature brought by the FVM. The adjacent mesh cells still share the same flux on a same face, and the flux leave from one cell will completely get into its neighbouring cell. The conservation of the proposed methods will also be checked in the coming Sec.\ref{sec::ChargeRelaxation} and Sec.\ref{sec::BubleRise}.

As for the overall solution procedure, the whole algorithm follows a time marching style and the solution procedure in each time step can be summarized as follows: The phase fraction is firstly updated with the algorithm described in Section \ref{sec::PhaseUpdate}. Then, the electric potential and charge distribution is obtained with the electrical physical properties obtained by the face discernment method and the convection flux obtained by the flux correction method. 
Subsequently, Eq.~(\ref{eq::disDisplacement})-(\ref{eq::disFeNew}) are solved to reconstruct the electric field strength and electric force. Finally, the PISO\cite{198640} algorithm is involved to solve the continuity equation and momentum equation with electric force and surface tension. The complete code is released as open source and obtainable from the GitHub warehouse\cite{GitHub} where more details of the algorithm can be found.

\section{Results and discussion\label{sec::ResultandDiscussion}}
In this section, five cases are presented to validate the performance of the face discernment and flux correction methods. Specifically, subsection \ref{sec::ChargeRelaxation} describes the charge relaxation under ohmic conduction to validate the accuracy and conservation of face discernment method; then, subsection \ref{sec::ChargeTube} considers the charge transportation within a tube to check the single-phase constriction feature of the flux-correction method; subsection \ref{sec::BubleRise} investigates the charge transportation within a three-dimensional bubble to confirm the conservation of flux-correction method; subsequently, subsection \ref{sec::dropletDeformation} simulates the droplet deformation under electric field to validate the algorithm performance of solving the coupled electric field and flow field; finally, subsection \ref{sec::taylorCone} performs a Taylor cone-jet case to check the ability of the algorithm to prevent the charge from leaking into insulating phase in real application scenarios.

\subsection{Charge relaxation through ohmic conduction \label{sec::ChargeRelaxation}}
\begin{figure}[h]
\centering
\includegraphics[scale=0.7]{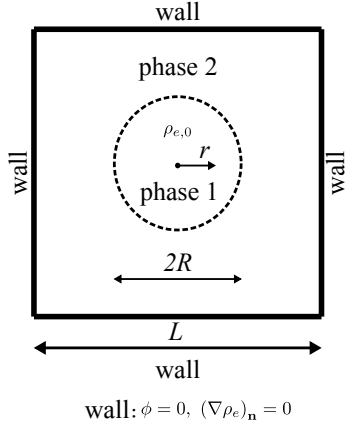}
\caption{\label{fig::cyc}The case configuration of the charge relaxation in a cylinder through ohmic conduction.}
\end{figure}

The sketch of the charge relaxation case\cite{Herrera} is illustrated in Fig.~\ref{fig::cyc}. There is a cylinder of radius $R$ located at a square domain of width $L$. The cylinder is a conducting medium with conductivity $K_1$ and permittivity $\varepsilon_1$ while the surrounding phase is insulating with $K_2$ and $\varepsilon_2$. Initially, the cylinder is set with an uniform charge density distribution of $\rho_{e,0}$. The charges will finally accumulate at the interface due to the
repulsion between them. The fluid motion is not considered and thus the charge is only transported by ohmic conduction.

\begin{figure}[h]
\centering
\subfigure{
	\label{fig::unrelaxed}
	\includegraphics[scale=0.4]{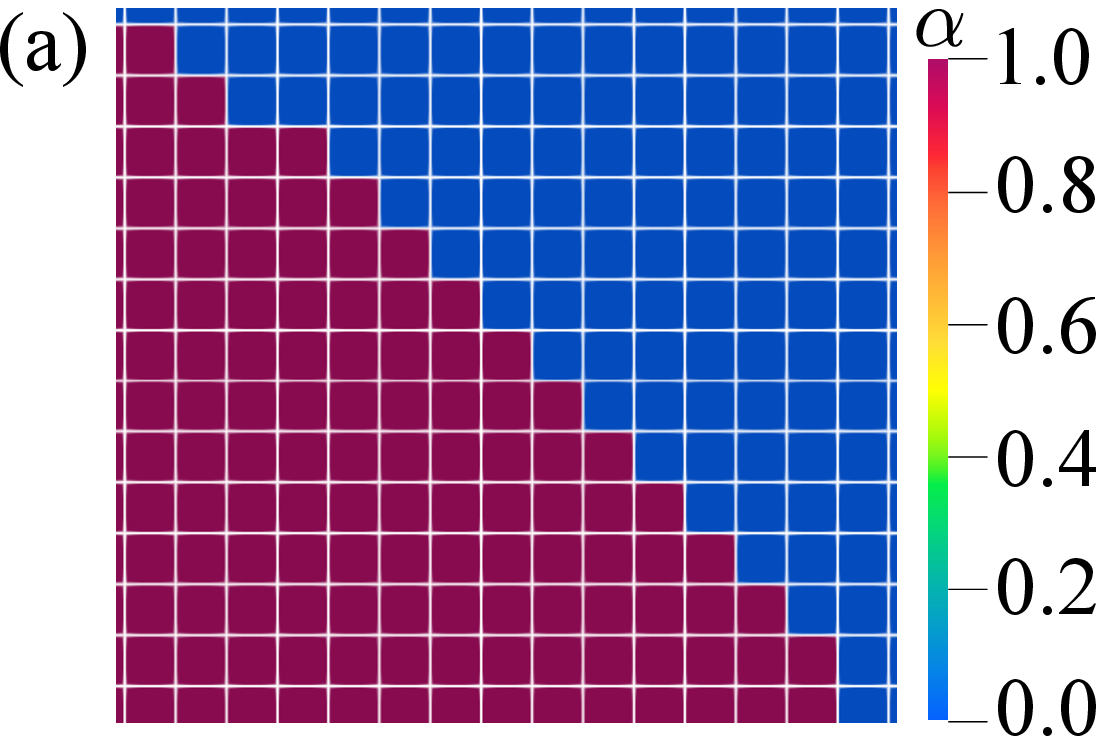}
}
\subfigure{
	\label{fig::relaxed}
	\includegraphics[scale=0.4]{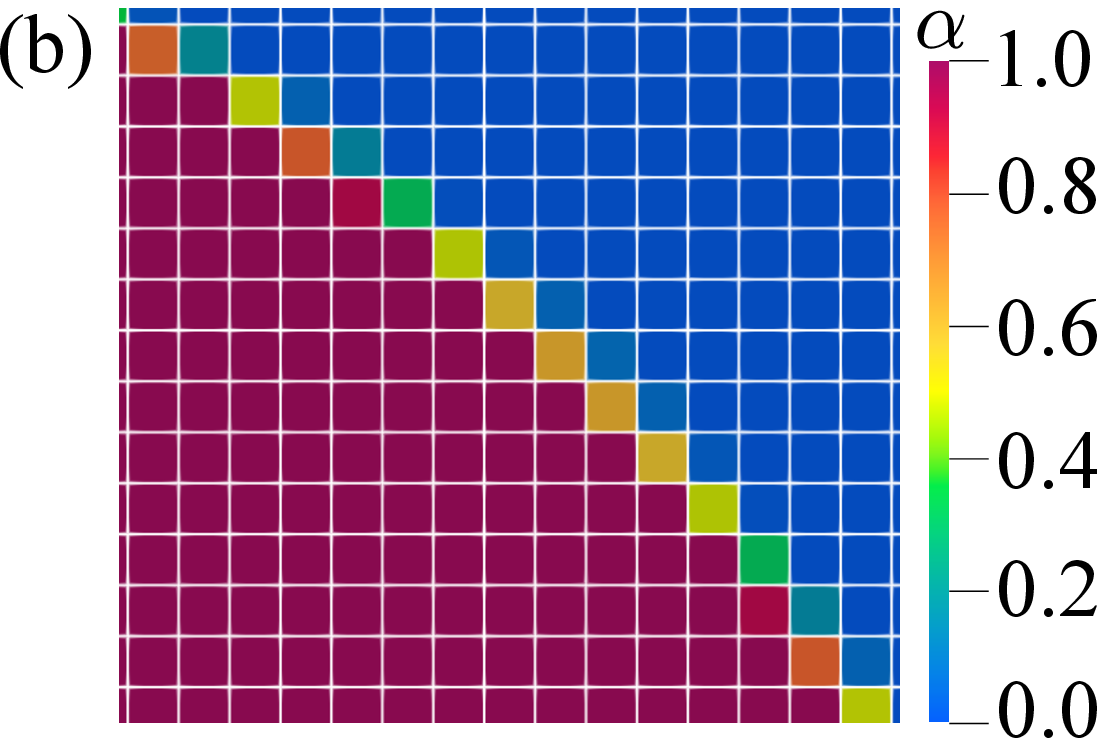}
}
\caption{\label{fig::relaxedAndUnrelaxed}Two different phase fraction distributions near the interface. (a) Unrelaxed distribution. (b) Relaxed distribution.
}
\end{figure}	

\begin{figure}[h]	
\centering
\subfigure{
	\label{fig::relaxation_rhoe_a}
	\includegraphics[scale=0.5]{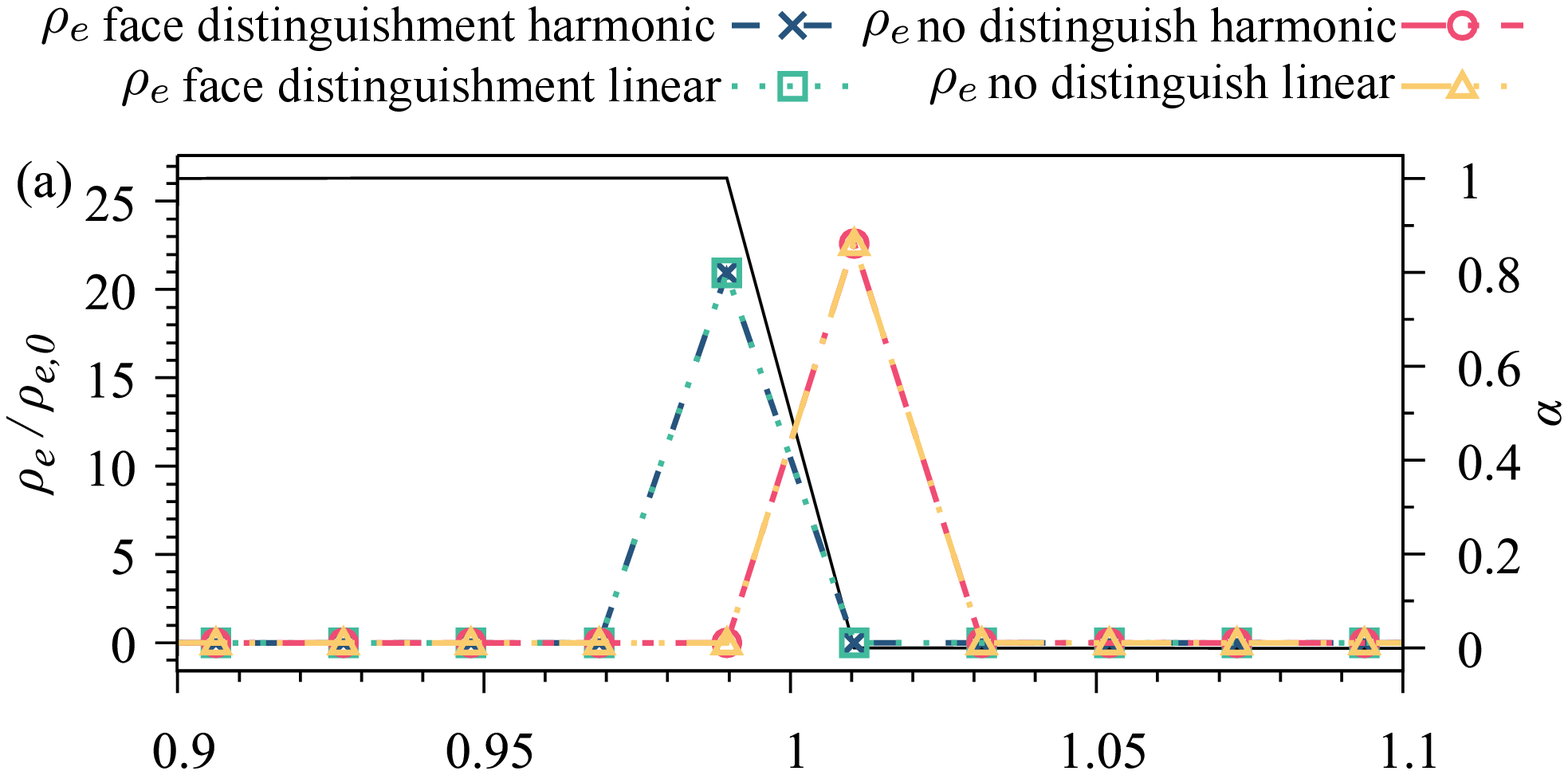}
}
\subfigure{
	\label{fig::relaxation_rhoe_b}
	\includegraphics[scale=0.5]{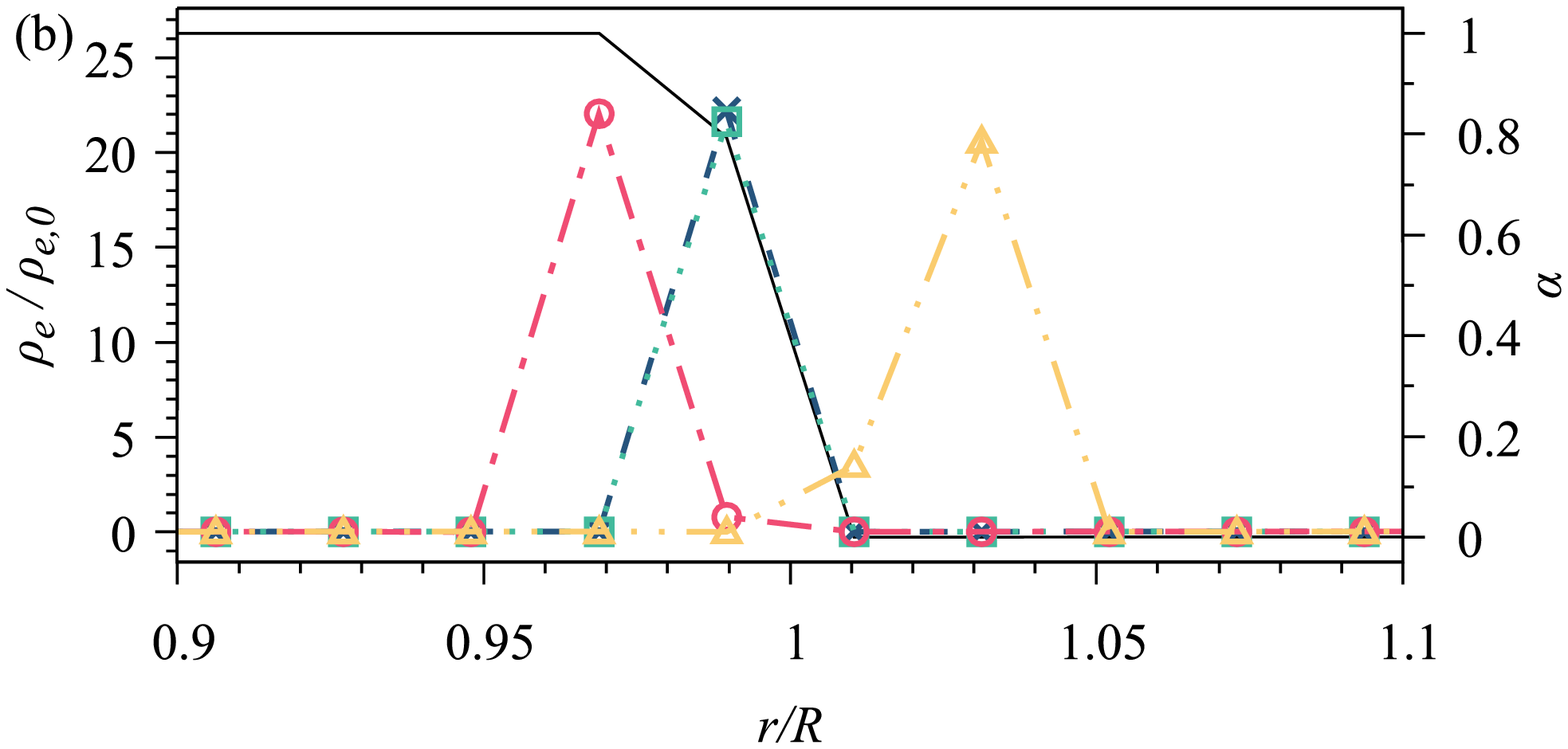}
}
\caption{\label{fig::relaxation_rhoe} The charge density distribution near the interface with different average methods for physical properties in the charge relaxation case when $t=30t_{e,1}$. The black solid line in the figure illustrates the phase fraction distribution. (a) Unrelaxed phase fraction field. (b) Relaxed phase fraction field.
}

\end{figure}

There are two kinds of phase distributions considered in this case. The first kind is called ``unrelaxed distribution'' as shown in Fig.~\ref{fig::unrelaxed} where the interface is situated at the cell faces, This kind of distribution often occurs as the initial field for phase fraction in simulations. The second distribution is ``relaxed distribution'' and it is illustrated in Fig.~\ref{fig::relaxed}. This distribution provides a smooth and diffusive phase fraction near the interface and it usually occurs during the simulation process.

In Fig.~\ref{fig::relaxation_rhoe}, the charge distribution near the interface is depicted when $t=30t_{e,1}$. Here, the $t_{e,1}$ represents the electric relaxation time $\varepsilon_1/K_1$ in phase 1. The parameters are set according to the previous study\cite{Herrera} as $L/R=20$, $K_2/K_1=0$, $\varepsilon_2/\varepsilon_1=2/3$ and the minimal mesh size is $R/48$. As shown in Fig.~\ref{fig::relaxation_rhoe_a}, the charge distribution is independent of the average method for physical properties when the phase fraction field is unrelaxed. The face discernment method can guarantee that the peak charge locates at the conducting cell, while the charge will accumulate at the insulating cell without the face discernment method. When the phase field is relaxed, the result obtained with the face discernment method is still independent of the physical property average method and the charge only appears at the interface cell whose $\alpha$ is between 0 and 1. However, for the situation where the face discernment method is not applied, the charge will be heaped up at the insulating cell when the linear average (Eq.~(\ref{eq::linearAverage})) is used and at the conducting cell when the harmonic average (Eq.~(\ref{eq::harmonicAverage})) is introduced. The charge in insulating phase obtained with the linear average method will move the peak Coulomb force from interface to insulating phase, leading to the generation of non physical velocity described in Sec.\ref{sec:introduction}.

\begin{figure}[htb]
\centering
\includegraphics[scale=0.5]{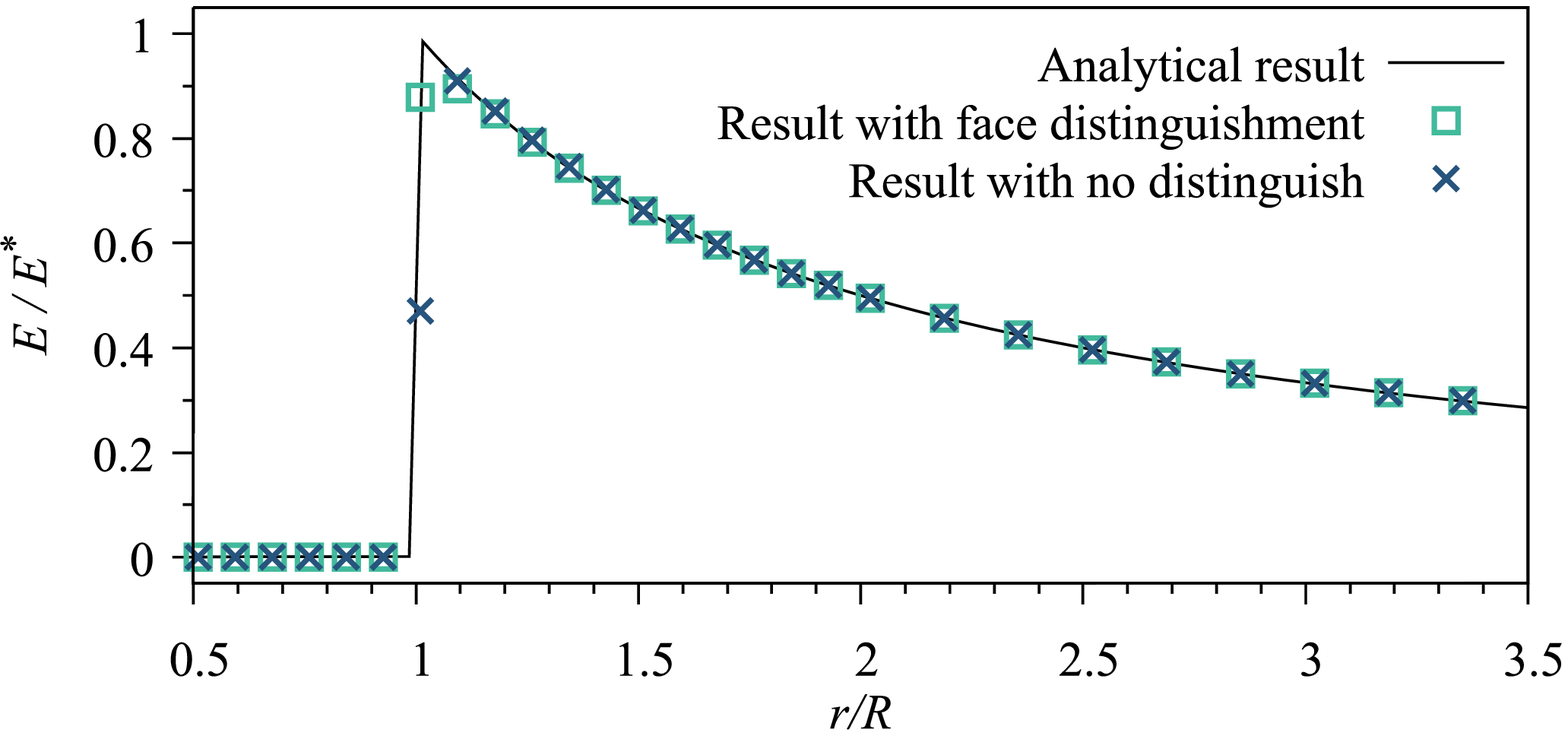}
\caption{\label{fig::relaxation_E}The electric field strength distribution near the interface with unrelaxed phase fraction field.}
\end{figure}

When the charge is fully relaxed, an analytical solution of electric field strength $E$ can be expressed as\cite{Herrera}:
\begin{eqnarray}
\frac{E(r/R)}{E^*} =
\left \{\begin{array}{lll}
	0 & r<R \\
	\frac{1}{r/R} & r\ge R
\end{array}\right.
\label{eq::relaxationE}
\end{eqnarray}
where $E^*=(R\rho_{e,0})/(2\varepsilon_2)$ is the characteristic electric field length. In Fig.~\ref{fig::relaxation_E}, the numerical distribution of electric field strength together with the analytical solution is presented and compared. The numerical results agree well with the analytical solution, and the Mean Squared Error(MSE) is 0.0324 and 0.0467 for the result with and without the face discernment method, respectively. The maximum MSE situates at the insulating cell nearest to the interface for both of the results and the magnitudes are 0.290 and 0.519 for face discernment and no distinguish case, respectively.

\begin{figure}[htb]
	\centering
	\includegraphics[scale=0.5]{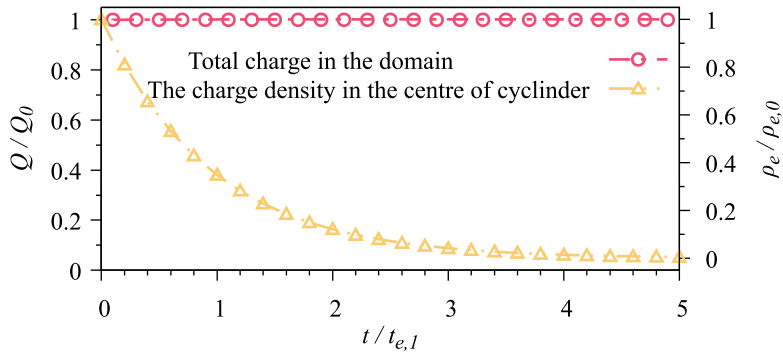}
	\caption{\label{fig::relaxation_conservation} The variation of total charge in the domain and the charge density in the centre of cylinder with time.}
\end{figure}

Theoretically, the total charge in the domain should remain unchanged during the relaxation process since the outside medium is insulating. Fig.~\ref{fig::relaxation_conservation} illustrates the variation of total charge $Q=\rho_e\pi R^2$ and the charge density at the centre of the cylinder with time. It is clear that the charge density at the cylinder centre declines gradually with time while the total charge is unchanged, which indicates that the face discernment method guarantees the conservation.

To conclude, the proposed face discernment method can successfully constrain the charge within the conducting phase and a higher accuracy is demonstrated than the no no distinguish cases. Besides, the implementation of the face discernment method does not wreck the conservation of the system. Unless otherwise specified, the face discernment method will always work in the cases shown in the coming sections and the average method is selected as harmonic.

\subsection{Charge transportation in tube\label{sec::ChargeTube}}

\begin{figure}[htbp]
\centering
\includegraphics[scale=0.7]{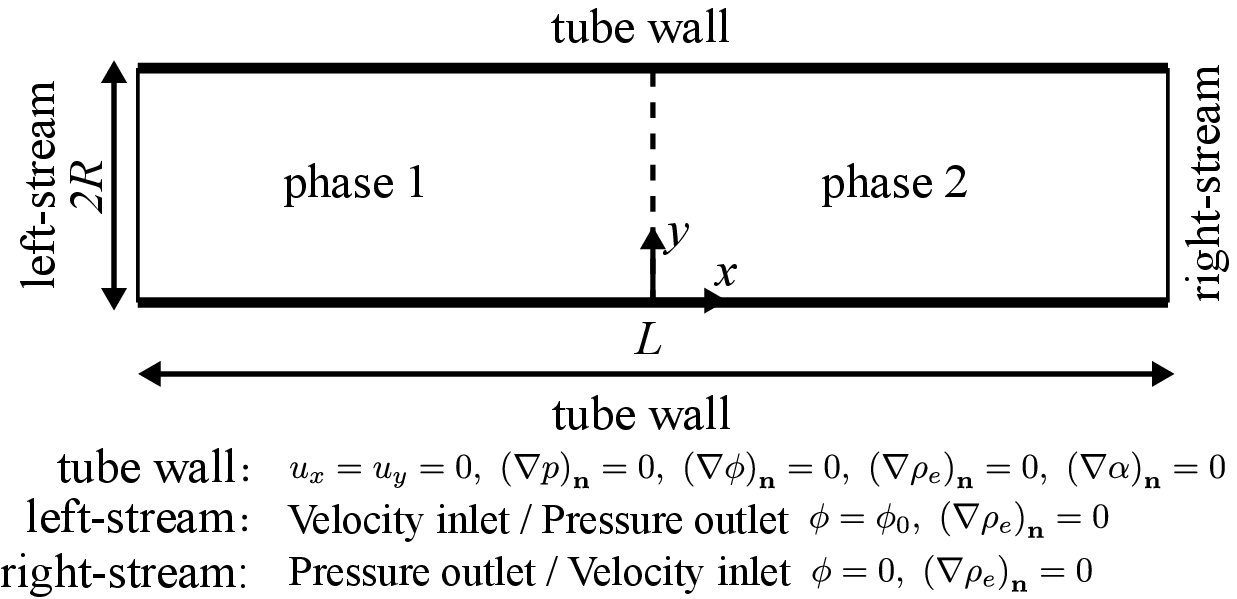}
\caption{\label{fig::tube}The case configuration of the charge transportation in a tube.}

\vspace{0.5cm}

\centering
\subfigure{
	\label{fig::sameCorrection}
	\includegraphics[scale=0.7]{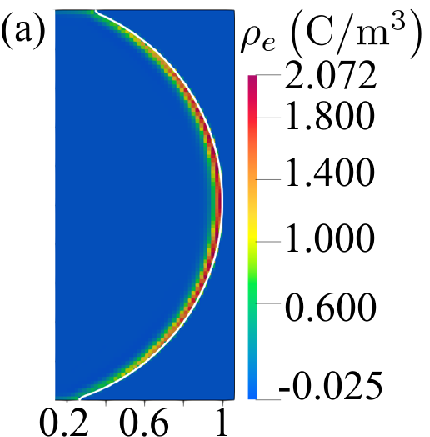}
}
\subfigure{
	\label{fig::sameAlpha}
	\includegraphics[scale=0.7]{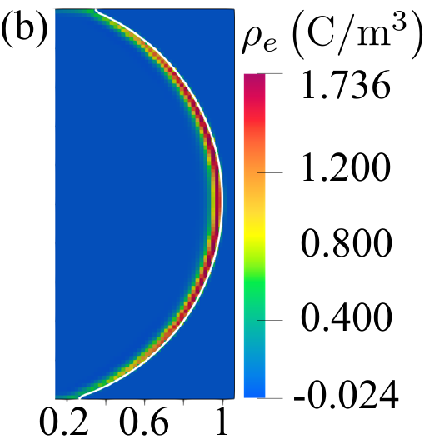}
}
\subfigure{
	\label{fig::sameNoCorrection}
	\includegraphics[scale=0.7]{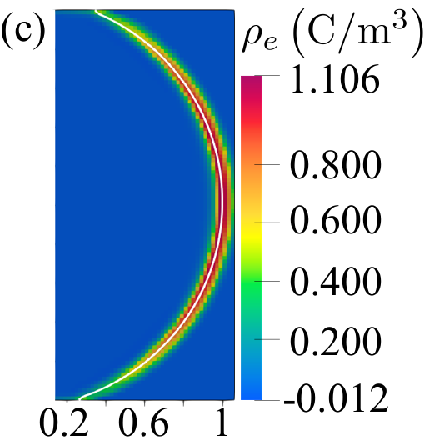}
}
\subfigure{
	\label{fig::differentCorrection}
	\includegraphics[scale=0.7]{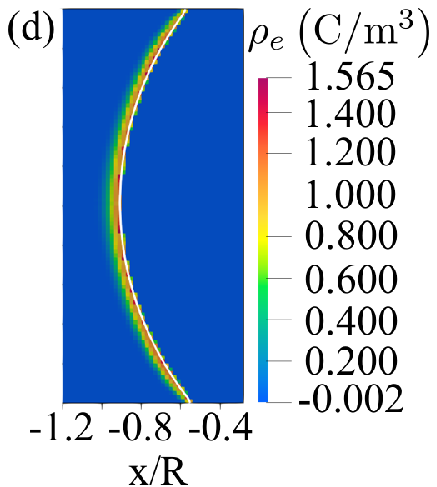}
}
\subfigure{
	\label{fig::differentAlpha}
	\includegraphics[scale=0.7]{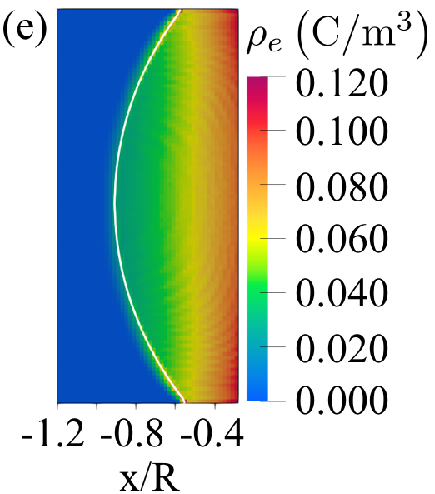}
}
\subfigure{
	\label{fig::differentNoCorrection}
	\includegraphics[scale=0.7]{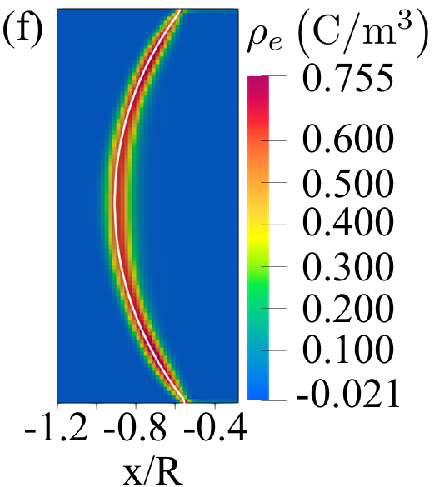}
}
\caption{\label{fig::chargeInterfaceCloud} The charge density distribution near the interface with different flux manipulation methods. The white dashed line illustrates the location of interface where $\alpha=0.5$. (a)-(c): The electric field and velocity are in the same direction. (d)-(e):The electric field and velocity are in the opposite direction. (a)(d): The result obtained with complete flux correction method. (b)(e): The result obtained only with single-phase correction step. (c)(f): The result obtained without any correction.
}
\end{figure}

\begin{figure}[tb]
\centering
\subfigure{
	\label{fig::transportRhoe_a}
	\includegraphics[scale=0.5]{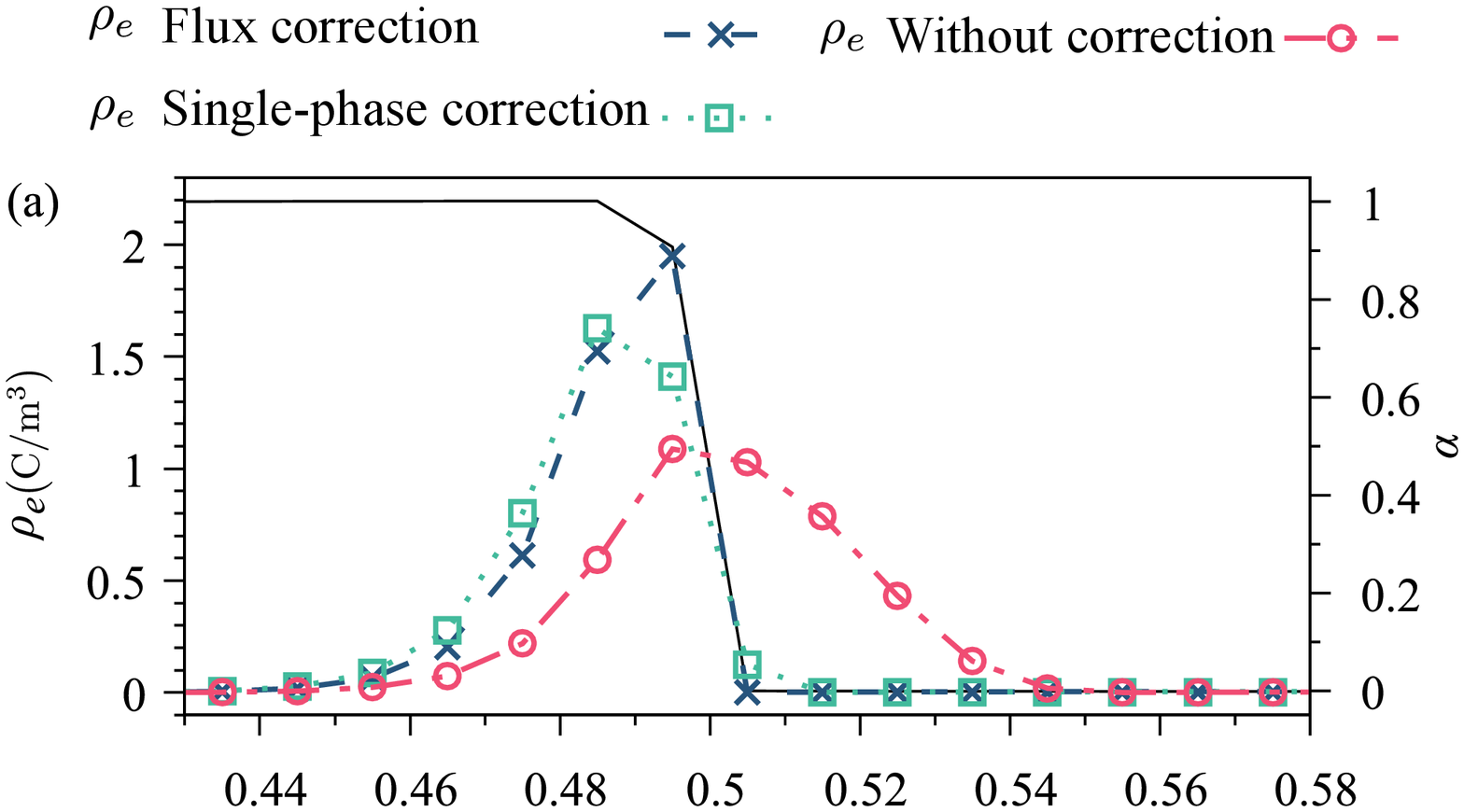}
}
\subfigure{
	\label{fig::transportRhoe_b}
	\includegraphics[scale=0.5]{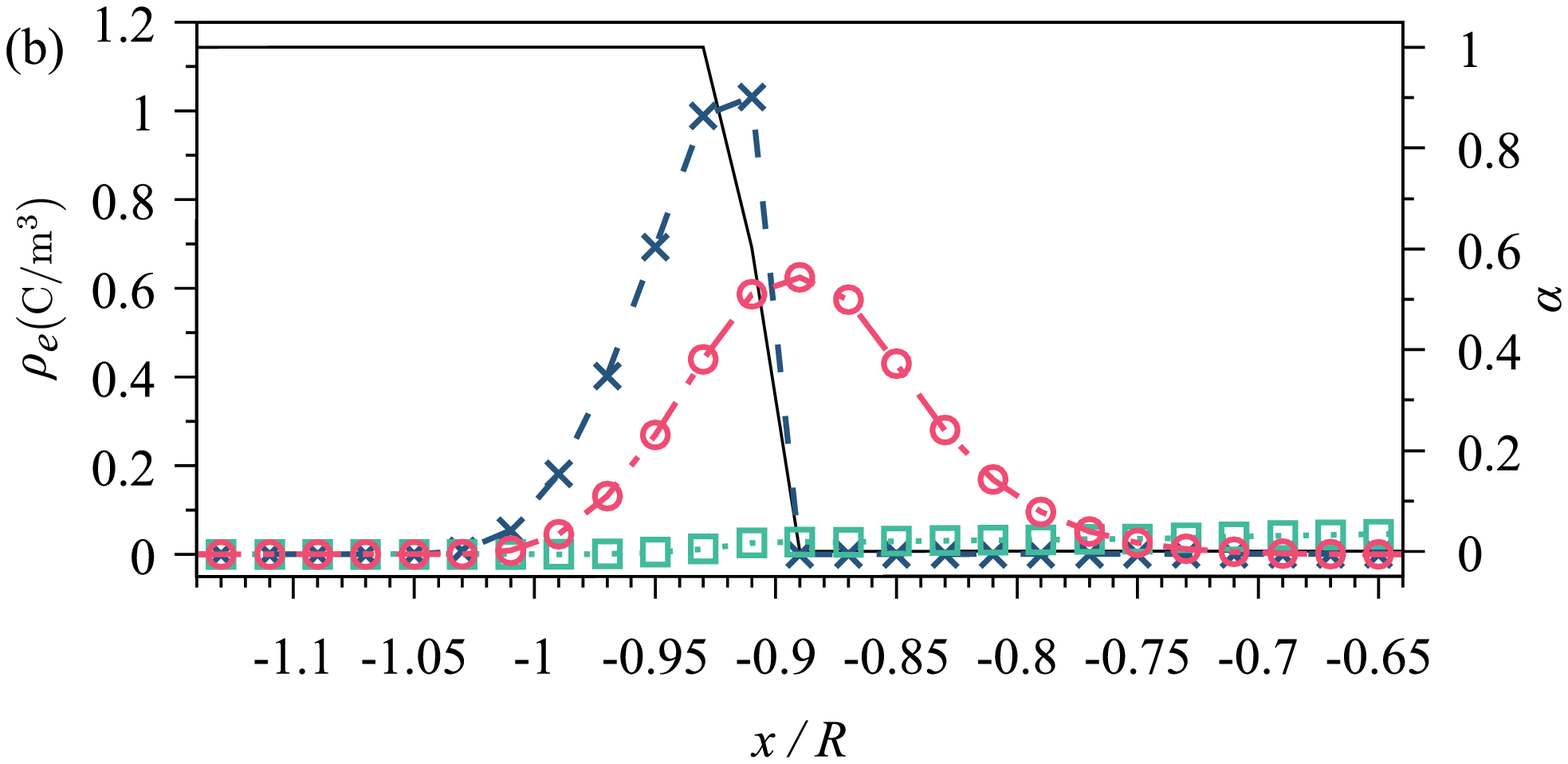}
}
\caption{\label{fig::transportRhoe} The charge density distribution at the line of $y=D/2$ with different flux manipulation method. The black solid line in the figure illustrates the phase fraction distribution. (a) The result when the electric field and velocity are in the same direction. (b) The result when the electric field and velocity are in the opposite direction.
}
\end{figure}
A case describing the charge transportation in a tube is designed in this section to check the single-phase constriction of the flux correction case. The simulation configuration is illustrated in Fig.~\ref{fig::tube}. There are two immiscible fluids located in the tube with their interface initially perpendicular to the side wall. The tube radius is $R$ and the length is $L$. The left side of the tube is imposed with a constant high potential $\phi_0$ while the right side is grounded. The flow inlet can be set either on the left or on the right. For the velocity inlet, a parabolic distribution is adopted as
\begin{equation}
u_x=\pm 2u_0\left ( 1-\frac{\left | y-R \right |^2 }{R^2} \right ) ,\ u_y=0
\label{eq::parabolic_velocity}
\end{equation}
Besides, the electric field and flow field is considered as unidirectional coupled in this case, that is, the electric force is absent in momentum equation.

The parameters used in the simulation are as follows. The radius and length of tube are set as $R=5$\textmu m and $L=40$\textmu m. The minimal cell size is set as $R/50$. The applied potential on the left side is $\phi_0=1\mathrm{V}$ and the velocity $u_0$ in Eq.~(\ref{eq::parabolic_velocity}) is $0.1\mathrm{m/s}$. Air (phase 1) and heptane (phase 2) are set as the working fluids and their physical properties are listed in Tab.\ref{tab::properties}.
\begin{table}[b]
\centering
\caption{Physical properties of fluids used in simulations.}
\label{tab::properties}
\begin{tabular*}{1\textwidth}{@{\extracolsep{\fill}}cccccc}
	\hline
	Fluid & $\rho\ \mathrm{(kg/m^3)}$ & $\mu\ \mathrm{(mPa\cdot s)}$ & $K\ \mathrm{(S/m)}$ & $\varepsilon\ \mathrm{(F/m)}$  & $\sigma\  \mathrm{(N/m)}$ \\
	\hline
	Air	& 1.225 & 0.018 & $1.050\times10^{-15}$ & $8.854\times10^{-12}$ &  \\
	Heptane \cite{Dastourani2021} & 684 & 0.420 & $1.4\times10^{-6}$ & $1.709\times10^{-12}$ & 0.019 \\
	1-octanol \cite{Herrera1} & 827 & 8.1 & $9\times10^{-7}$ & $8.854\times10^{-11}$ & 0.027 \\
	\hline
\end{tabular*}
\end{table}

In Fig.\ref{fig::chargeInterfaceCloud} and Fig.\ref{fig::transportRhoe}, the charge density distribution near the interface is depicted. Here, besides the complete flux correction method, the results obtained only with the single-phase correction flux is also shown to highlight the role of the additional correction flux in flux correction method. What's more, the correction flux is directly obtained from the phase fraction without the face discernment process in the results of pure single-phase correction method to investigate the role of face discernment method in the simulation. As shown in Fig.\ref{fig::sameCorrection}-\ref{fig::sameNoCorrection} and Fig.\ref{fig::transportRhoe_a}, both the flux correction method and pure single-phase correction step can successfully limit the charge density in the heptane phase when the flow directs from heptane to air. However, the result without any correction shows a very diffusive charge layer at the interface. The flux correction also leads to a sharper charge peak compared with the pure single-phase correction step thanks to the contribution made by the face discernment method. When the flow velocity directs from air to heptane, as indicated in Fig.\ref{fig::differentCorrection}-\ref{fig::differentNoCorrection} and Fig.\ref{fig::transportRhoe_b}, the charge density in the result of flux correction method is still only situated in the heptane region and the distribution in the no correction case is diffusive as before. However, the charge is almost fully stuck in the air region when only the single-phase correction step is applied, which is due to the lack of convection as highlighted in Fig.~\ref{fig::transportSketch}.

To sum up, the flux correction method is capable of guaranteeing the single-phase transportation of charge when the velocity acting on the interface is from any direction, and the additional correction flux in flux correction method plays a critical role in preventing the charge from stucking in the insulating fluid when the flow directs to the conducting phase.
\subsection{Charge transportation in a bubble\label{sec::BubleRise}}

\begin{figure}[htb]
\centering
\includegraphics[scale=0.7]{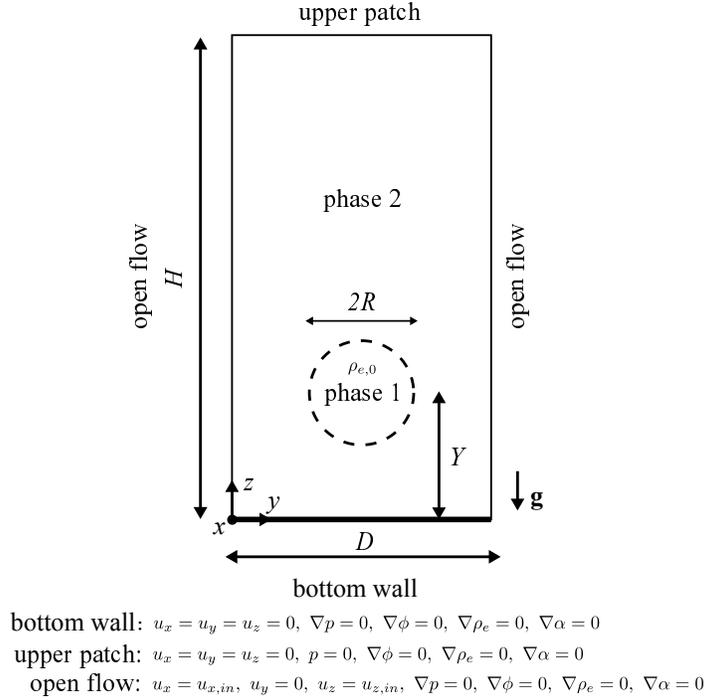}
\caption{\label{fig::bubble}The case configuration of the charge transportation in a bubble. The $u_{x,in}$ and $u_{y,in}$ in the open flow boundary refer to the velocity in the cells adjacent to the boundary.}
\end{figure}
\begin{figure}[htb]
	\centering
	\subfigure{
		\label{fig::chargeInbuble_a}
		\includegraphics[scale=0.5]{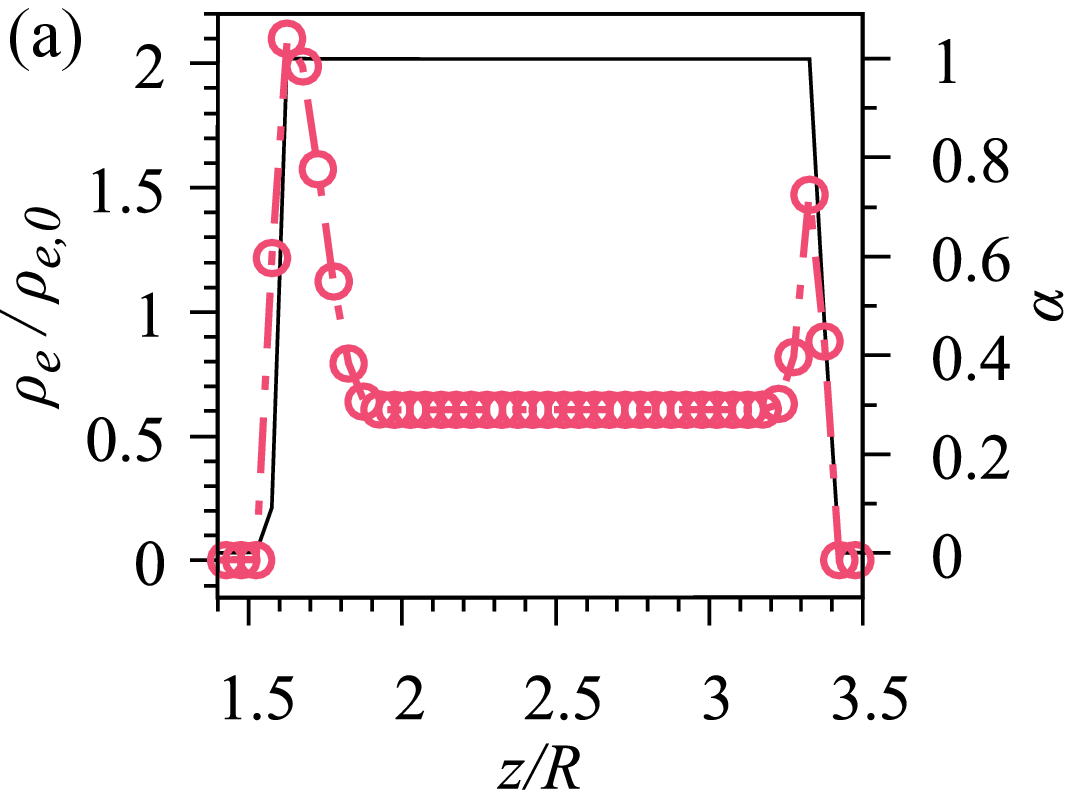}
	}
	\subfigure{
		\label{fig::chargeInbuble_b}
		\includegraphics[scale=0.5]{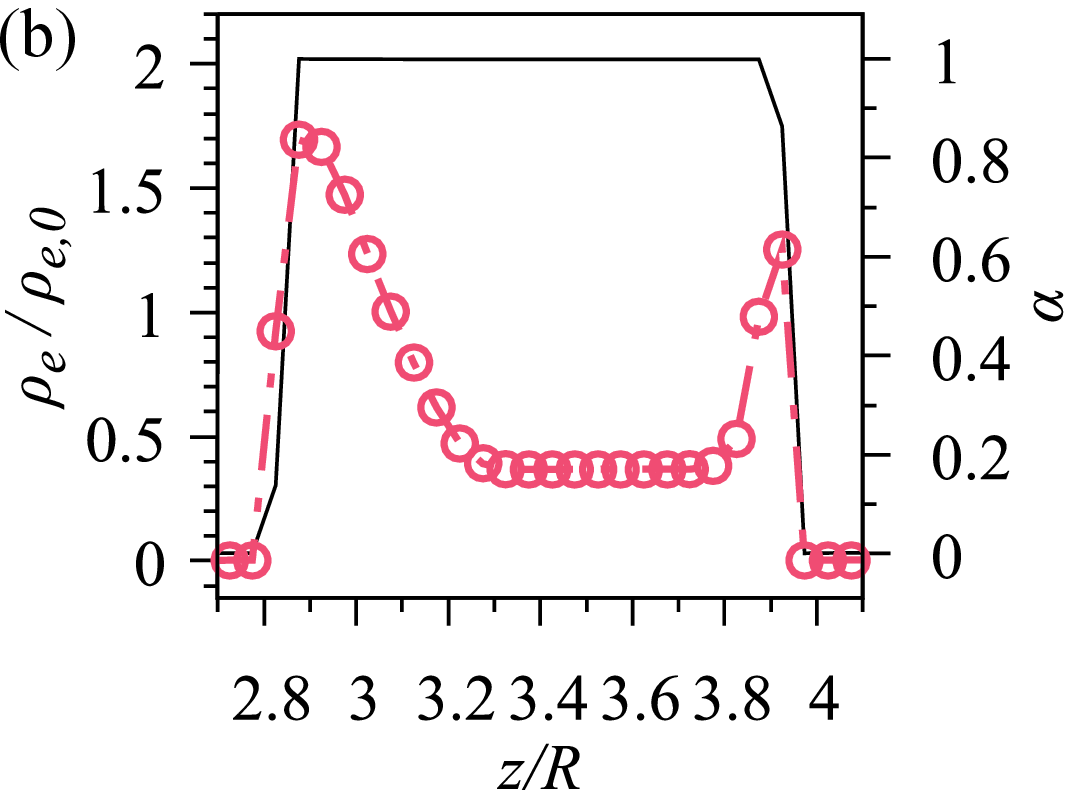}
	}
	\subfigure{
		\label{fig::chargeInbuble_c}
		\includegraphics[scale=0.5]{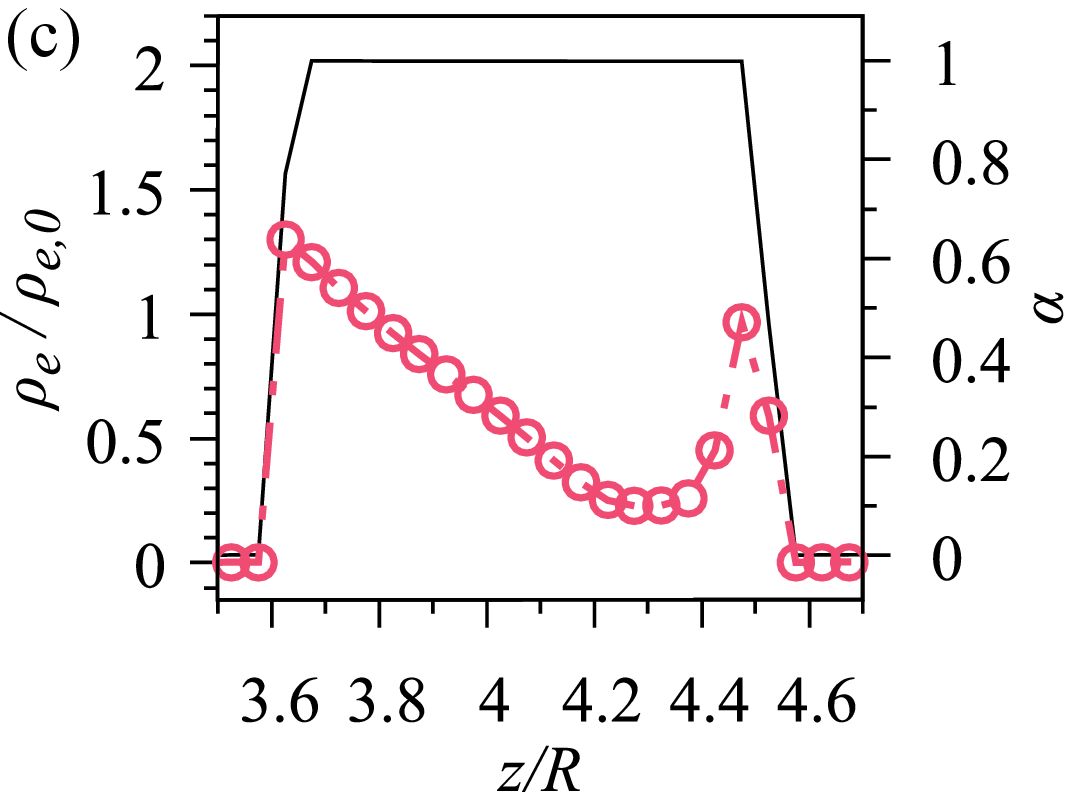}
	}
	\subfigure{
		\label{fig::chargeInbuble_d}
		\includegraphics[scale=0.5]{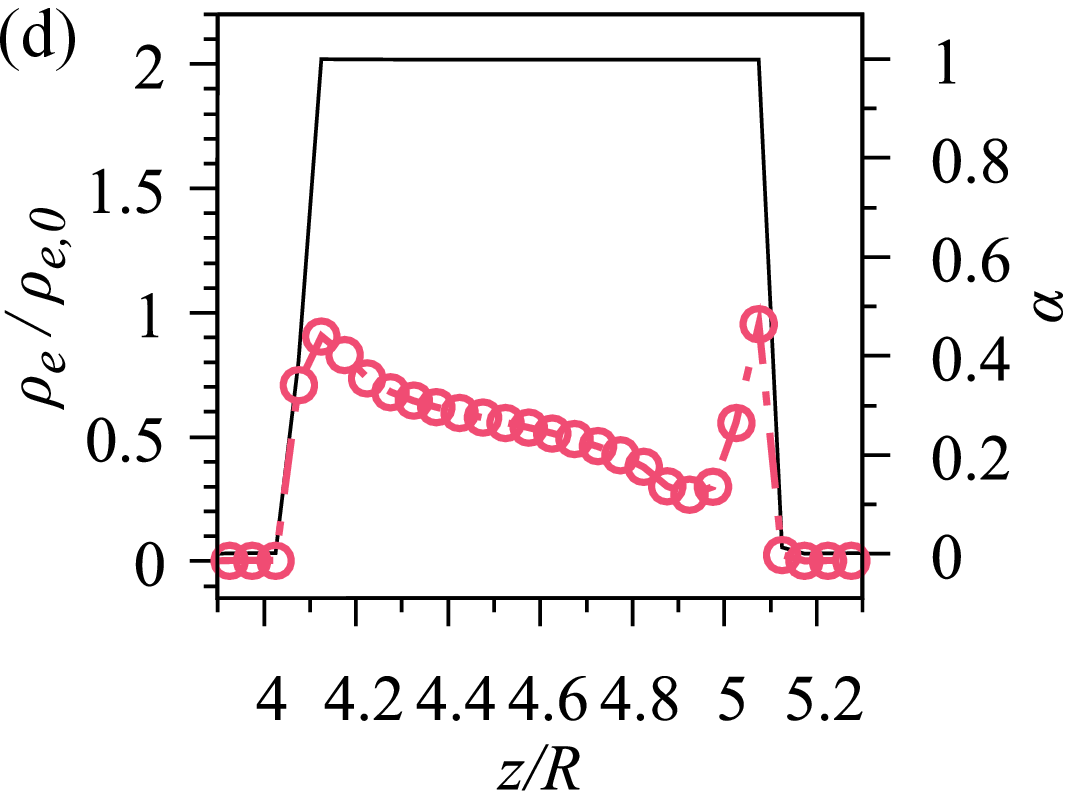}
	}
	\caption{\label{fig::chargeInbuble} The charge density and phase fraction distribution in the central axis of the cylinder at different time moments. The black solid line represents the phase fraction distribution while the red dashed line with circle shows the charge density. (a) $t=0.5t_{e,1}$. (b) $t=1.0t_{e,1}$. (c) $t=1.5t_{e,1}$. (d) $t=2.0t_{e,1}$.
	}
	
\end{figure}
\begin{figure}[htbp]
\centering
\subfigure{
	\label{fig::bubble_a}
	\includegraphics[scale=0.4]{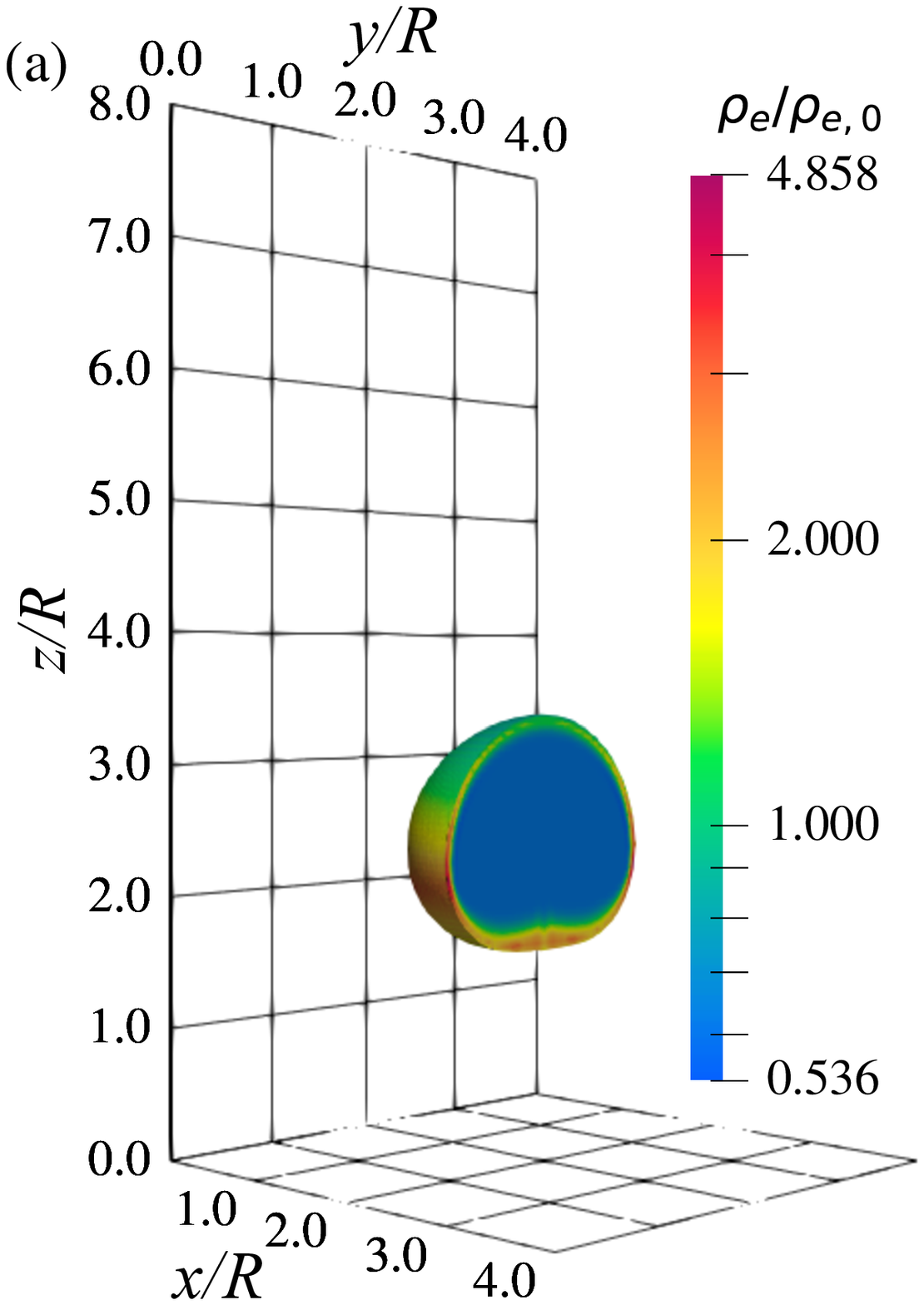}
}
\hspace{0.5cm}
\subfigure{
	\label{fig::bubble_b}
	\includegraphics[scale=0.4]{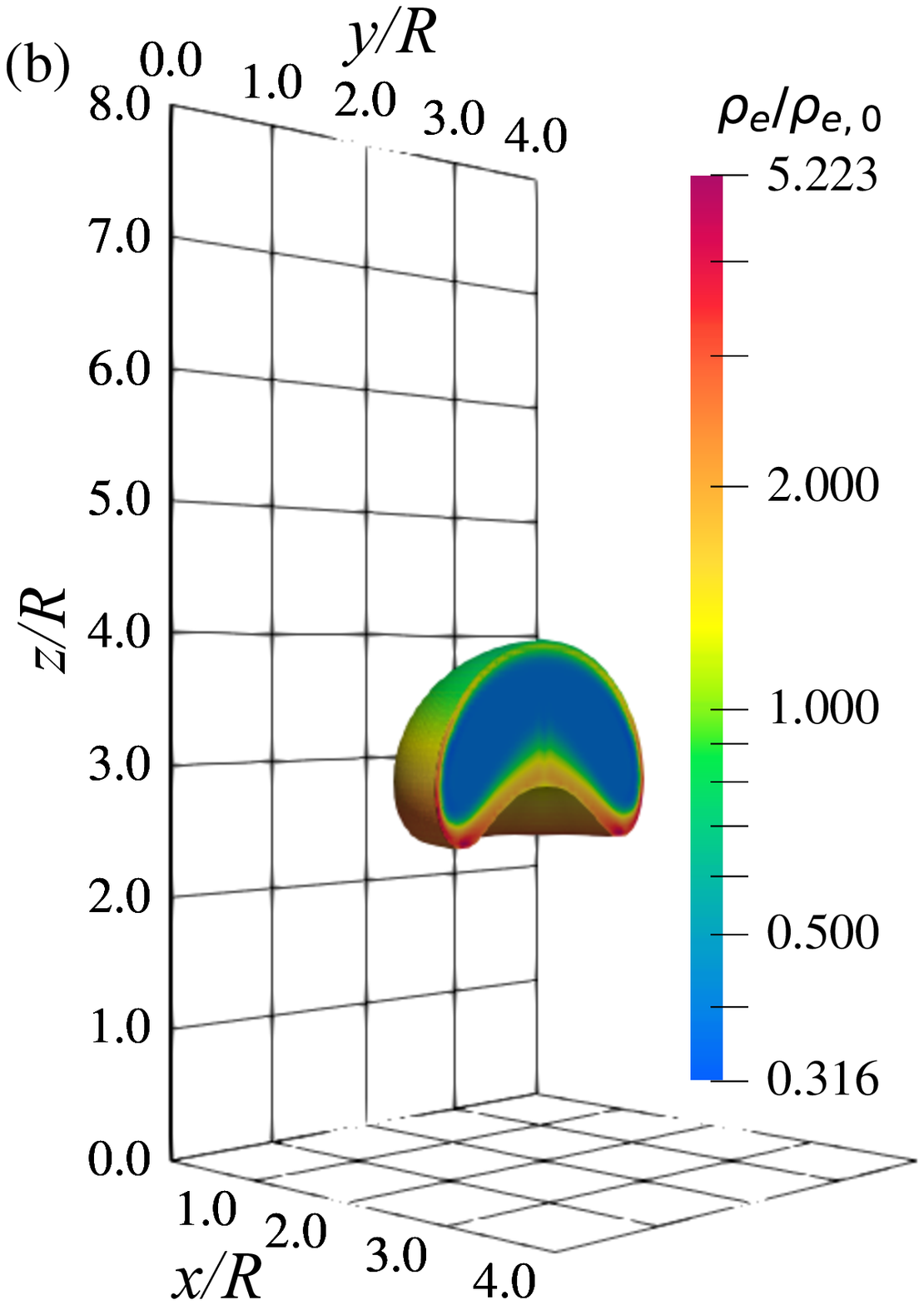}
}\\
\subfigure{
	\label{fig::bubble_c}
	\includegraphics[scale=0.4]{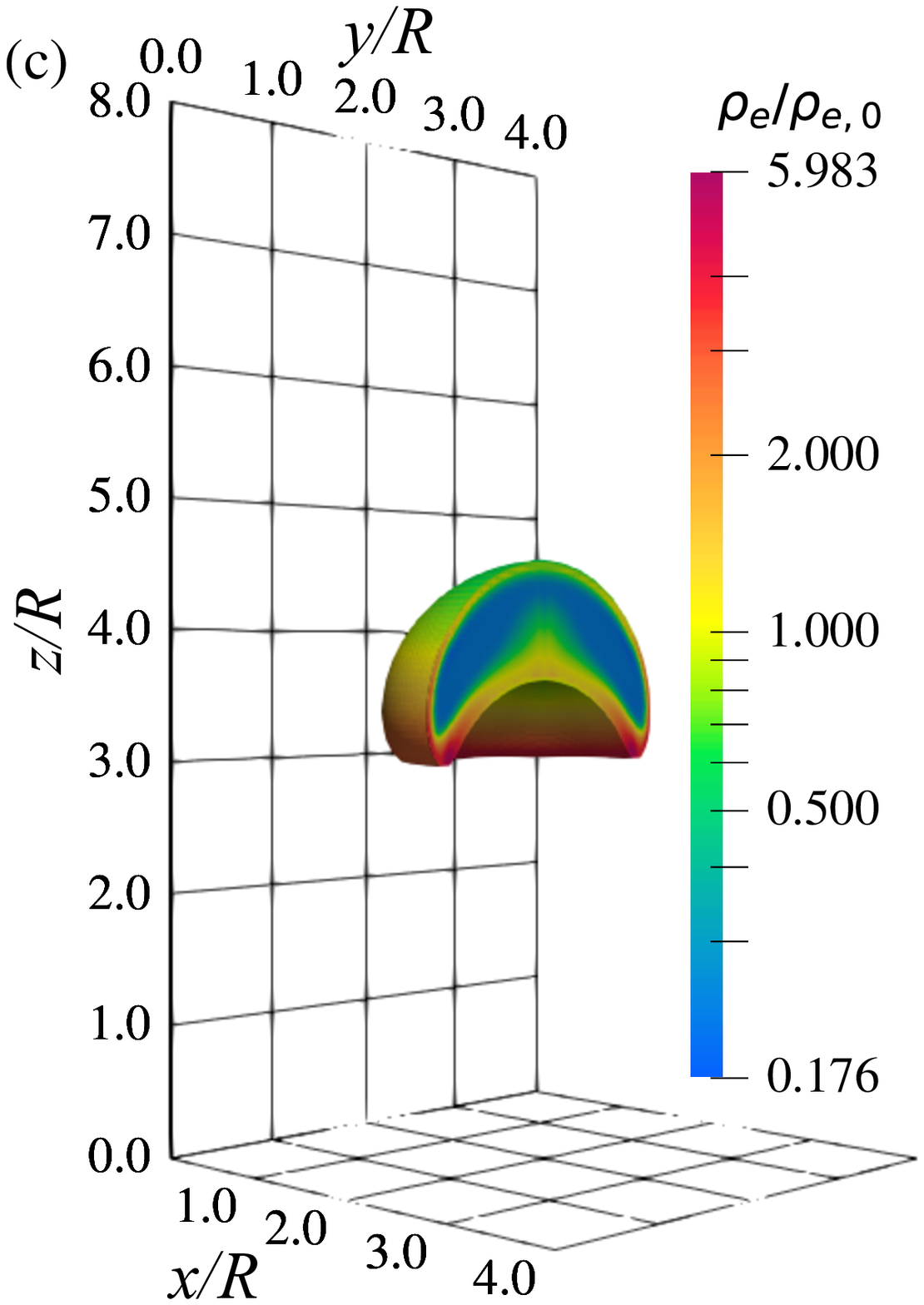}
}
\hspace{0.5cm}
\subfigure{
	\label{fig::bubble_d}
	\includegraphics[scale=0.4]{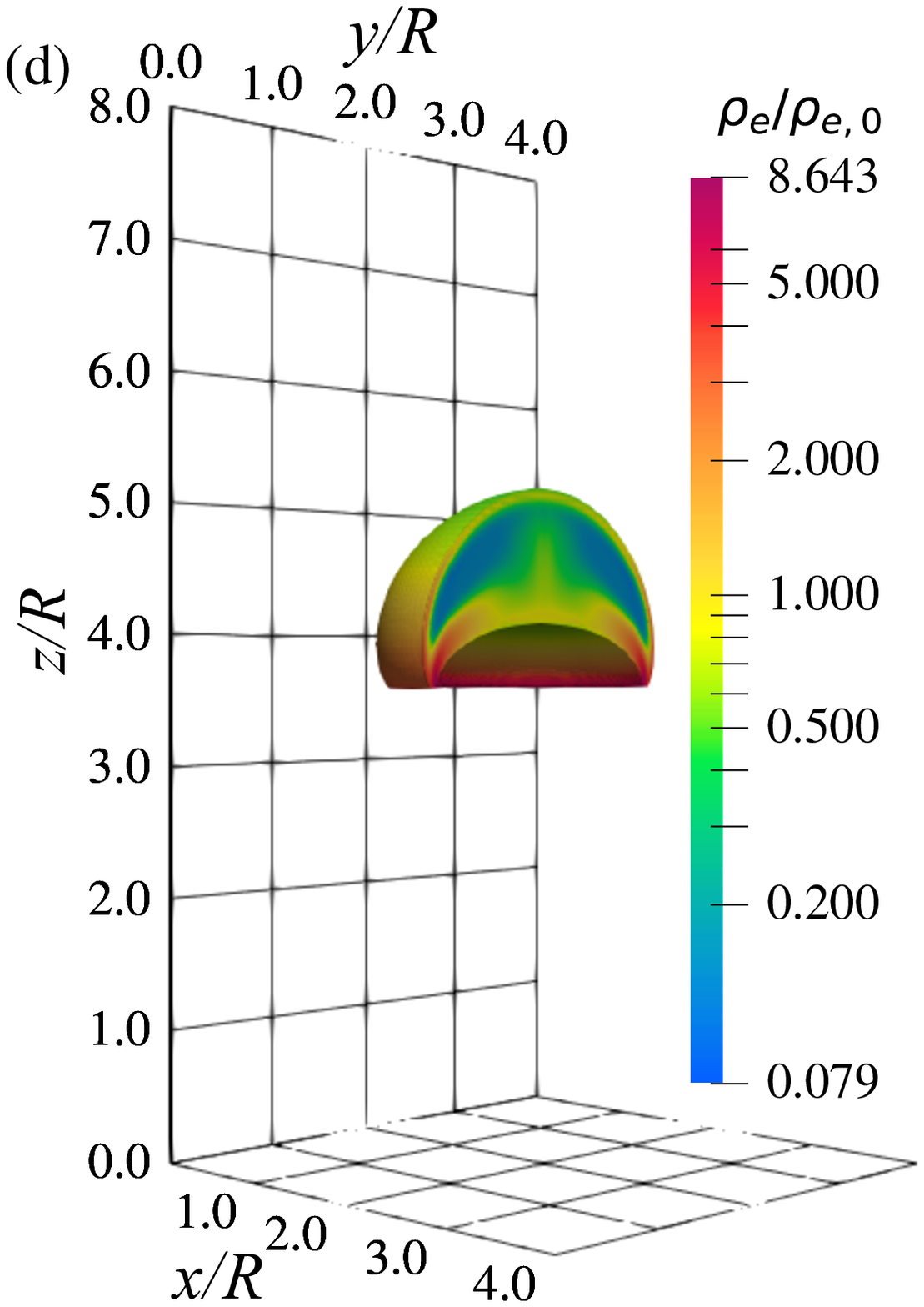}
}
\caption{\label{fig::bubbleRaise} The charge density distribution in the rising bubble at different time moments. (a) $t=0.5t_{e,1}$. (b) $t=1.0t_{e,1}$. (c) $t=1.5t_{e,1}$. (d) $t=2.0t_{e,1}$.
}

\end{figure}

\begin{figure}[htbp]
\centering
\includegraphics[scale=0.5]{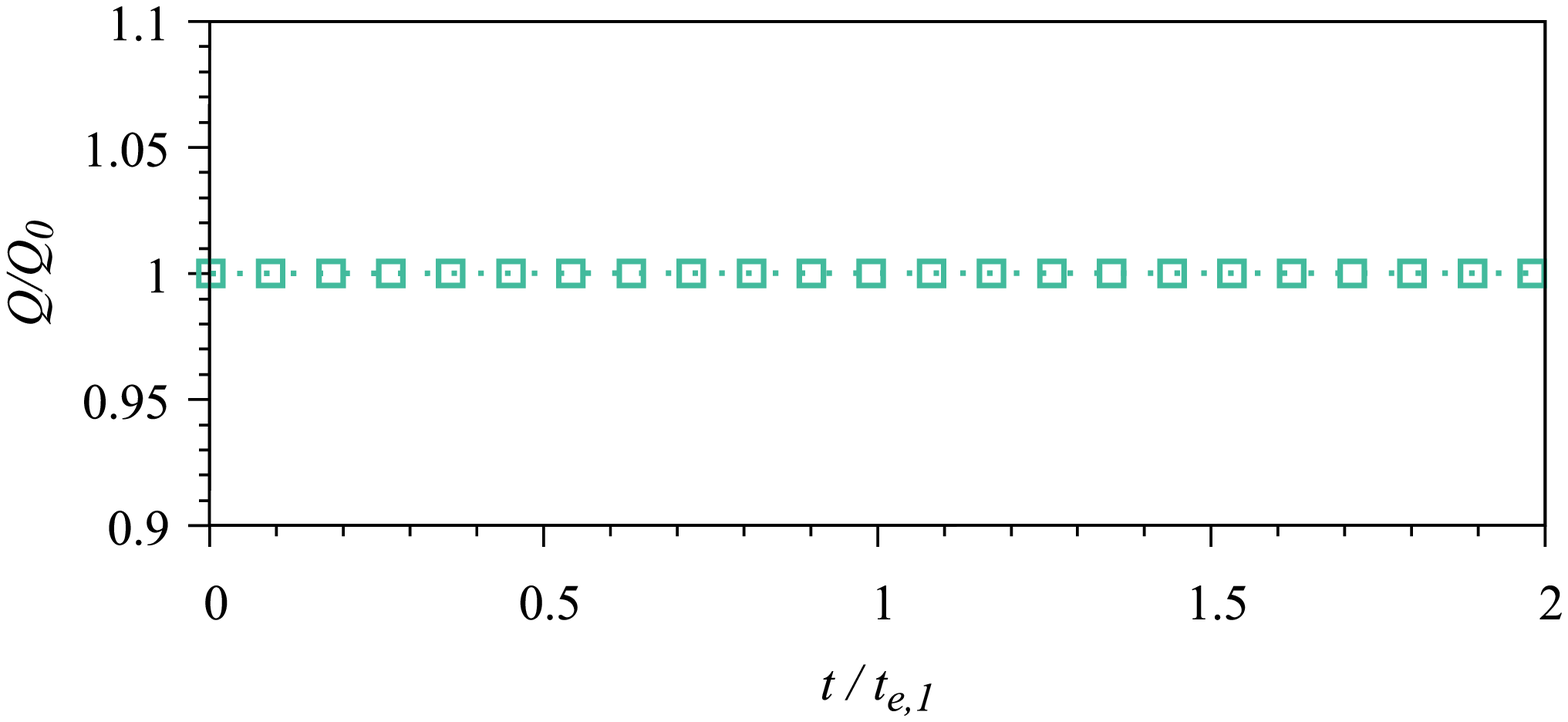}
\caption{\label{fig::fluxConservation} The variation of total charge with time in the bubble rising case.}
\end{figure}
In this subsection, a test case considering the charge transportation in a bubble is designed to verify the conservation of flux-correction method and the sketch of this case is drawed in Fig.~\ref{fig::bubble}. A three-dimensional bubble with radius $R$ is suspended in the outer fluid filled in a cylinder domain with diameter $D$ and height $H$. The bubble is initially padded with a uniform charge with the density of $\rho_{e,0}.$ Under the action of buoyancy, the bubble will slowly float up and deform.  The charge in the bubble will get relaxed due to both the convection and ohmic conduction. Also, the electric field and flow field are considered as unidirectional coupled and the electric force is absent in momentum equation.

The single rising bubble case is a classical test case for two-phase flow which has a well-documented simulation parameters sets as\cite{Quantitative2009,3DDrop,Gamet2020}: $D=4R$, $H=8R$, $Y=2R$,  $\rho_1/\rho_2=0.001$, $\mu_1/\mu_2=0.01$, Bond number $Bo=\rho_2 g (2R)^2/\sigma=125$ and Galilei number $Ga=\rho_2g^{1/2}(2R)^{3/2}/\mu_2=35$. The electric parameters are set as subsection \ref{sec::ChargeRelaxation} and the ratio of capillary time\cite{Herrera} of flow field and charge relaxation time is set as $\left[\rho_2\left(2R\right)^3/\sigma\right]^{1/2}/t_{e,1}=8$. The minimal mesh cell is set as $R/32$.

The distribution of charge is illustrated in Fig.\ref{fig::chargeInbuble} and Fig.\ref{fig::bubbleRaise}. Before the significant deformation and movement of the bubble, the charge distribution within the bubble is dominated by the ohmic conduction and the charge tends to migrate from the inside of the bubble to the interface. When the bubble begins to move upward under the action of buoyancy, there is a depression in the lower part of the bubble and the charge will accumulate at the lower tip edge. The charge transportation on the upper side of the bubble is now similar to the case in Fig.\ref{fig::sameCorrection}-\ref{fig::sameNoCorrection} while the bottom side resembles the situation in Fig.\ref{fig::differentCorrection}-\ref{fig::differentNoCorrection}. With the growth of deformation, the convection at the bulge of the lower interface becomes gradually greater than local ohmic conduction, so that the charge in the bubble is transported from the lower side to the upper side in the centre axis of bubble (Fig.\ref{fig::chargeInbuble_b}-\ref{fig::chargeInbuble_d}), forming a plume like charge distribution in the bubble (Fig.\ref{fig::bubble_b}-\ref{fig::bubble_d}). Fig.\ref{fig::chargeInbuble} also indicates that the charge only exists at the conducting and interface cell, which means that the face discernment and flux correction method work well with three-dimensional geometry. Besides, the variation of total charge amount in the simulation domain is drawn in Fig.\ref{fig::fluxConservation} and a good conservation is guaranteed by the flux correction method according to this figure.

\subsection{Droplet deformation under electric field\label{sec::dropletDeformation}}

The electric force is absent in the previous subsections to validate the passive transport of free charge under ohmic conduction and convection. Starting from this subsection, the electric force together with the two-way coupling between the electric field and the flow field will be included. First, the droplet deformation under electric field is discussed in this section.

\begin{figure}[h]
\centering
\includegraphics[scale=0.7]{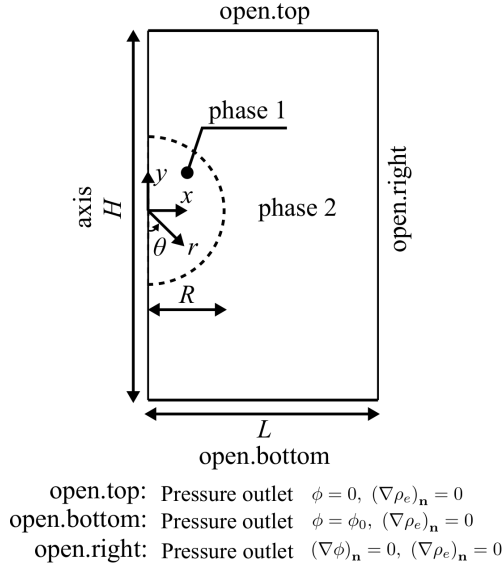}
\caption{\label{fig::droplet}The case configuration of the droplet deformation under electric field.}
\end{figure}

As illustrated in Fig.~\ref{fig::droplet}, a droplet with radius $R$ is immersed in a second medium. An electric field of $E_\infty=\phi_0/H$ is applied to the droplet. After the simulation starts, the positive and negative charges will migrate to both ends of the droplet under the action of electric field. Finally, the droplets will deform under the action of Coulomb force and dielectric force. Taylor once provided the analytical solution for the deformation as\cite{Herrera,Tomar2007,Taylor1966}
\begin{equation}
A=\frac{9}{16}
\frac{C a_{E}}{(2+B)^{2}}
\left[
1+B^{2}-2 Q+\frac{3}{5}(B-Q) \frac{2+3 \lambda}{1+\lambda}
\right]
\label{eq::deformatioAnalytical}
\end{equation}
where $B=K_1/K_2$, $Q=\varepsilon_1/\varepsilon_2$ and $\lambda=\mu_1/\mu_2$. $C a_{E}=E_{\infty}^{2} R_{d} \varepsilon_{2} / \sigma$ is the electric capillary number and $A$ is a function validating the deformation amplitude:
\begin{equation}
A=\frac{b-a}{b+a}
\
.
\label{eq::deformatioFunction}
\end{equation}
Here, $b$ and $a$ are the droplet length after deformation in the $y$ direction (parallel to the electric field) and $x$ direction (perpendicular to the electric field) respectively.

The ratio of conductivity $B$ ranging from 2 to 13 is set as an independent variable to obtain different deformation amplitude. Since there is no insulating fluid, the flux correction method will not be introduced while the face discernment method is still used to obtain the sharp distribution of physical properties. Other parameters are set as $Q=10$, $\lambda=1$, $C a_{E}=0.18$. The simulation domain is configured as axisymmetric with $L=4R$ and $H=8R$. The minimal mesh cell size is $R/80$.
\begin{figure}[h]
\centering
\includegraphics[scale=0.5]{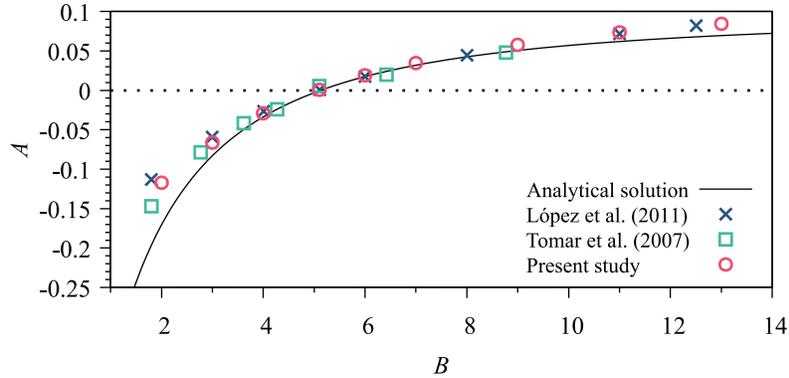}
\caption{\label{fig::deformationValue} The droplet deformation amplitude with different conductivity ratio $B$.}
\end{figure}

Fig.\ref{fig::deformationValue} depicts the simulated deformation amplitude with different conductivity ratio. When $B<5.1$, the deformation is perpendicular to the direction of the electric field and thus the function $A$ is negative. By comparison, the deformation is parallel to the electric field when $B>5.1$. The obtained deformation amplitude is compared with the analytical solution of Eq.~(\ref{eq::deformatioAnalytical}) and other numerical results\cite{Herrera,Tomar2007}. In the small deformation stage of the droplet, say $4\le B \le 5$, the numerical solution is in good agreement with both the analytical solution and the literature results. However, there is an obvious difference between the numerical solution and the analytical solution regardless of the deformation direction when the droplet deformation increases. The literature results also deviate from the analytical solution. This may due to that Eq.~(\ref{eq::deformatioAnalytical}) is obtained through the linear theory and the error may be amplified in the large deformation region due to the nonlinear effect.

\begin{figure}[htbp]
\centering
\subfigure{
	\label{fig::deformationVelocity}
	\includegraphics[scale=0.4]{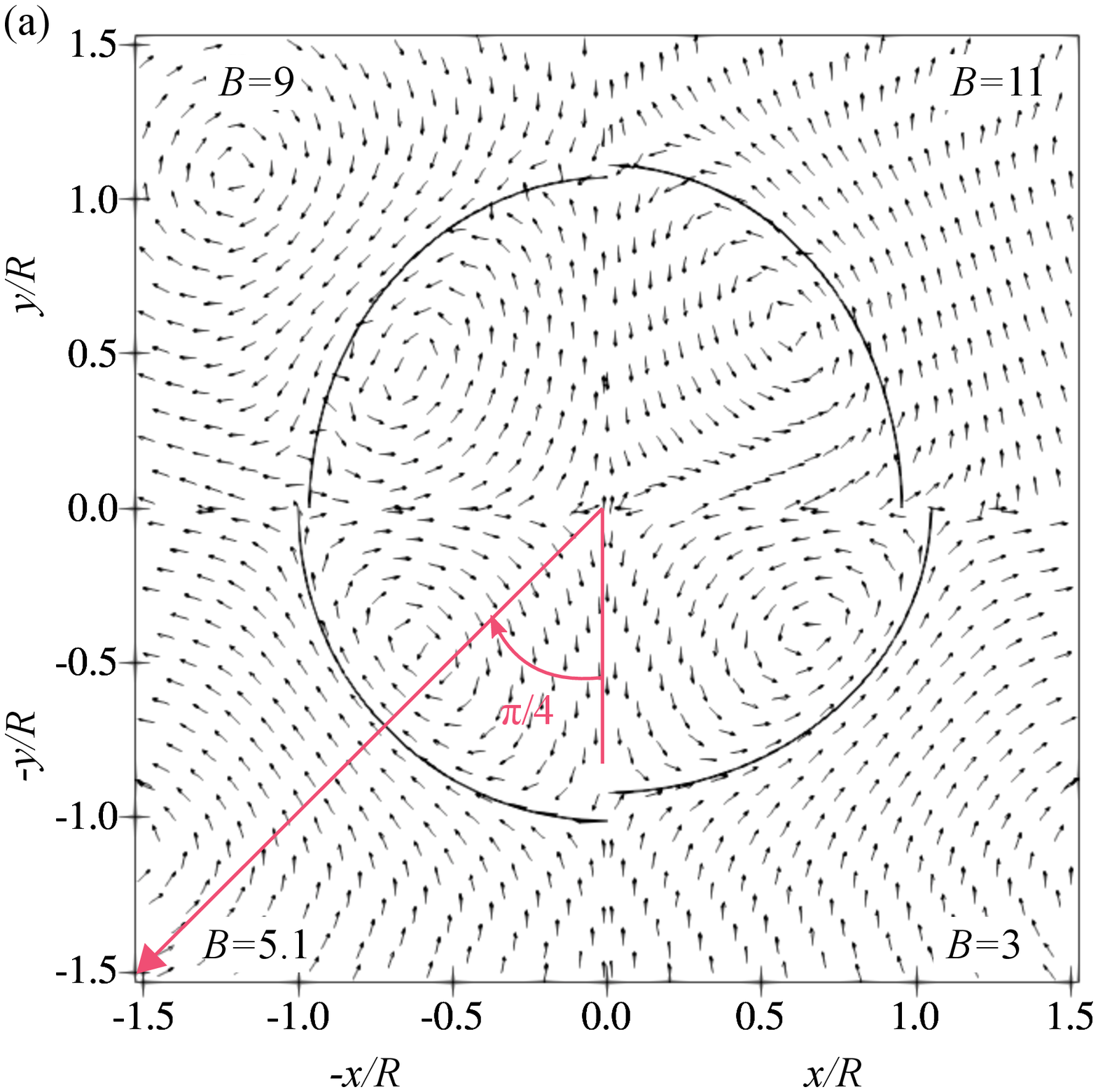}
}
\subfigure{
	\label{fig::deformationCharge}
	\includegraphics[scale=0.4]{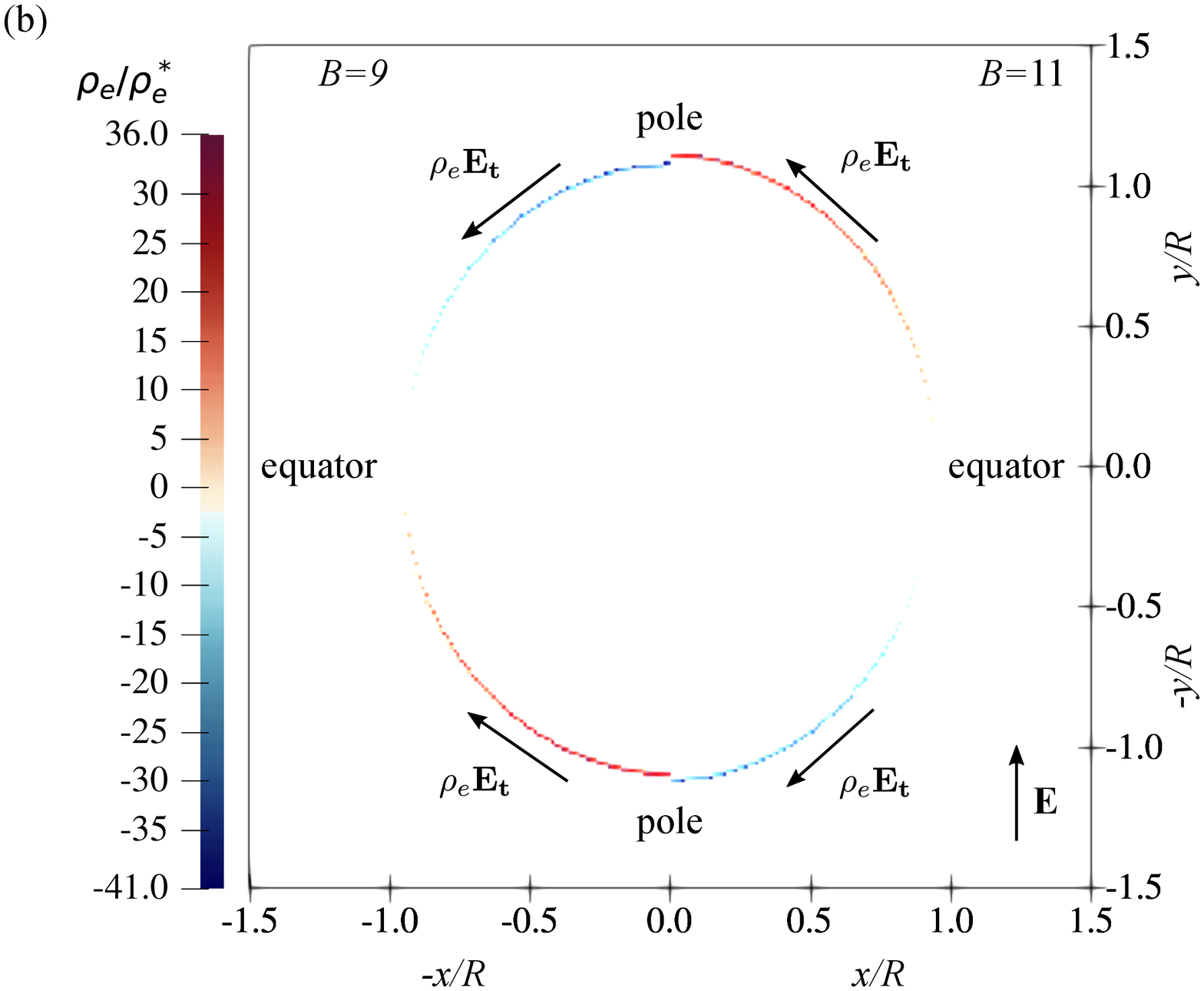}
}
\caption{\label{fig::deformationVelocityAndCharge} The droplet velocity and charge density distributions with different $B$. The characteristic charge density $\rho_{e}^*$ is chosen as $\varepsilon_{1}\phi_0/D^2$ (a) The velocity distribution when $B=3$, $B=5.1$, $B=9$ and $B=11$. The black solid line represents the interface. (b) The charge density distribution when $B=9$ and $B=11$.}
\end{figure}

\begin{figure}[t]
\centering
\includegraphics[scale=0.5]{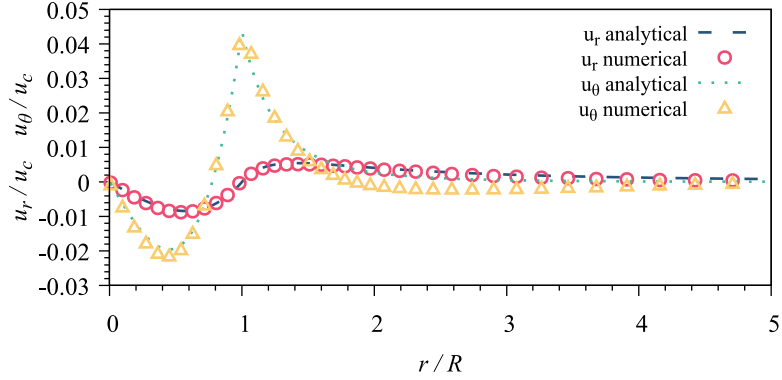}
\caption{\label{fig::dropVelcoityLine} The velocity distribution of the droplet when $B=5.1$.}
\end{figure}

The droplet outline and velocity distributions with different $B$ are shown in Fig.~\ref{fig::deformationVelocity}. The vortex flow near the interface, which is considered as one of the characteristics of the multiphase EHD flow, can be observed clearly. As indicated in Eq.~(\ref{eq::interStressT})-(\ref{eq::maxwellN}), the dielectric force only works in the normal direction of the interface while the Coulomb force $\rho_{e} \mathbf{E}$ owns a tangential component at the interface. This tangential force can only be balanced by the viscous stress induced by fluid flow, and it eventually results in the vortex flow near the interface. Taylor pointed out that the direction of the vortex is from pole to equator at the interface when $B/Q < 1$, and from equator to pole when $B/Q > 1$\cite{Taylor1966}. This is because the electric relaxation time of the outer media $t_{e,2}=\varepsilon_2/K_2$ is smaller than the droplet electric relaxation time $t_{e,1}=\varepsilon_1/K_1$ when $B/Q < 1$, that is, the charge of the outer liquid reaches the interface faster than that of the inner droplet. So the charge at the interface mainly comes from the outer medium. Similarly, the free charge accumulated at the interface mainly comes from the droplet when when $B/Q > 1$. Fig.~\ref{fig::deformationCharge} intuitively shows the change of charge polarity at the interface when $B/Q > 1$ and $B/Q < 1$. The reversal of charge polarity brings about the reversal of Coulomb force, finally resulting in the change of vortex direction shown in Fig.~\ref{fig::deformationVelocity}. The analytical solution\cite{Herrera,Tomar2007,Taylor1966} for this velocity distribution in polar coordinate can be written as
\begin{eqnarray}
\frac{u_r}{u_c} =
\left \{\begin{array}{lll}
	N \frac{r}{R}\left[1-{\left(\frac{r}{R}\right)}^{2}\right]\left(3 \cos ^{2} \theta-1\right)
	&
	r<R
	\\
	N\left[{\left(\frac{r}{R}\right)}^{-4}-{\left(\frac{r}{R}\right)}^{-2}\right]\left(3 \cos ^{2} \theta-1\right)
	&
	r\ge R
\end{array}\right.
\label{eq::ur}
\end{eqnarray}
and
\begin{eqnarray}
\frac{u_\theta}{u_c} =
\left \{\begin{array}{lll}
	\frac{3 N}{2} \frac{r}{R}\left(1-\frac{5}{3} {\left(\frac{r}{R}\right)}^{2}\right) \sin 2 \theta
	&
	r<R
	\\
	-N {\left(\frac{r}{R}\right)}^{-4} \sin 2 \theta
	&
	r\ge R
\end{array}\right.
\label{eq::ut}
\end{eqnarray}
where
\begin{equation}
N=\frac{9}{10} \frac{1}{(1+\lambda)} \frac{B-Q}{(B+2)^{2}}
\label{eq::N}
\end{equation}
and
\begin{equation}
u_c= \frac{R \varepsilon_{2} E_{\infty}^{2}}{\mu_{2}}
\
.
\end{equation}
Note that the coefficient is 9/10 in Eq.~(\ref{eq::N}) rather than $-9/10$ appearing in the expression (45) of L\'opez-Herrera et.al \cite{Herrera}. The former one is consistent with the earliest analytical solution of Taylor\cite{Taylor1966} as well as the numerical result of Tomar\cite{Tomar2007}. 
Fig.\ref{fig::dropVelcoityLine} depicted the velocity distribution obtained by the present algorithm on the line of $\theta=\pi/4$ when $B=5.1$ compared with the analytical solution. The numerical solution agrees well with the analytical one, which indicates a good performance of the proposed algorithm.

\subsection{Taylor cone-jet\label{sec::taylorCone}}
\begin{figure}[htb]
\centering
\includegraphics[scale=0.7]{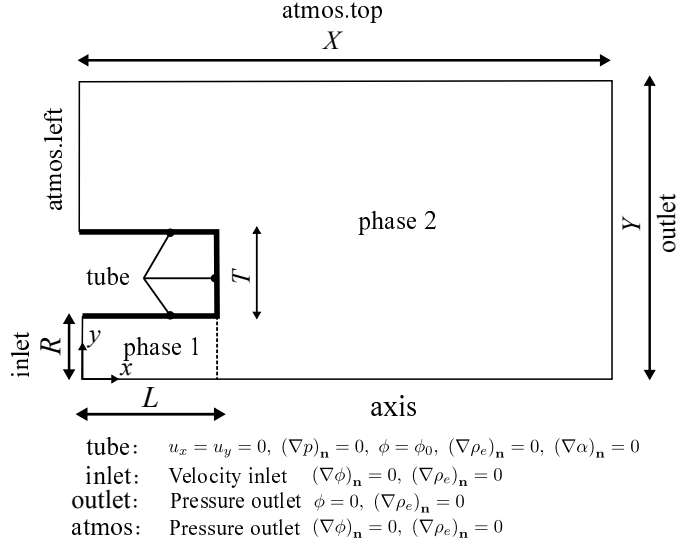}
\caption{\label{fig::conejet} The case configuration of the Taylor cone-jet. }
\end{figure}
In this subsection, a Taylor cone-jet case is designed to highlight the superiority of the proposed method in a real physical phenomena involving a perfectly insulating phase. The sketch of this case is depicted in Fig.~\ref{fig::conejet}. The working medium is initially filled in the capillary with radius $R$ and the tube is imposed with a high potential $\phi_0$. Under the action of external electric field, the fluid in the capillary will be sucked out and formulates a cone-jet with the balance of viscous force, surface tension and electric force.

The parameters are set according to the experimental configuration of Herrera et.al\cite{Herrera1}. The fluid inside the tube and the surrounding medium is set as 1-octanol and air, respectively, and  their properties are listed in the Tab.~\ref{tab::properties}. The radius and thickness of the capillary is 0.1 mm and 0.01 mm respectively. The grounded electrode is 1 mm away from the tube outlet. The height of the domain $Y$ and the length of tube $L$ in the simulation domain are set as 0.7 mm and 0.15 mm respectively. The potential $\phi_0$ is set as 1600 V and the velocity is set according to the given flow rate of $Q=$1mL/h with a parabolic distribution described in Eq.~(\ref{eq::parabolic_velocity}). The minimal mesh cell size is set as $R/40$.

\begin{figure}[htb]
\centering
\includegraphics[scale=0.4]{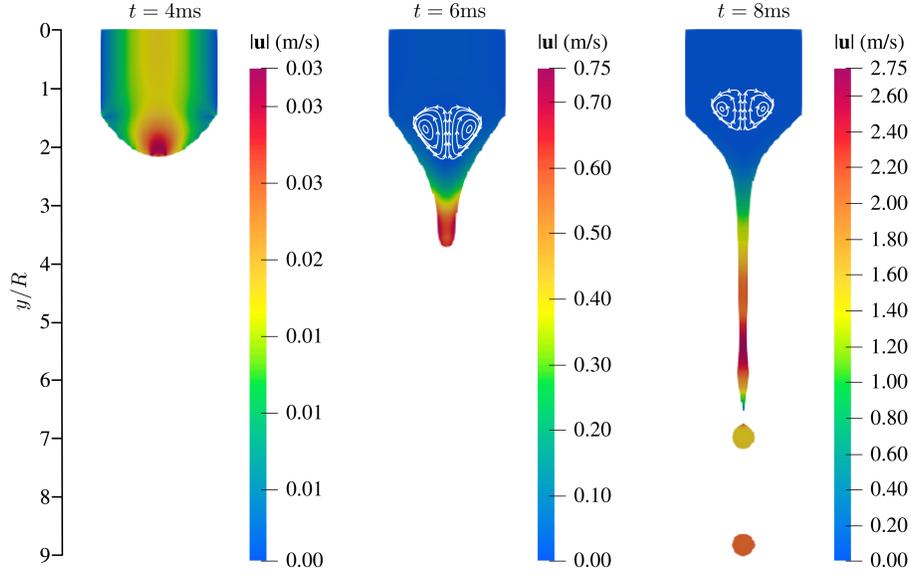}
\caption{\label{fig::jetProcess} The jet shape and velocity distribution inside the one jet during the jet formation process. }
\end{figure}

\begin{figure}[htbp]
\centering
\includegraphics[scale=0.5]{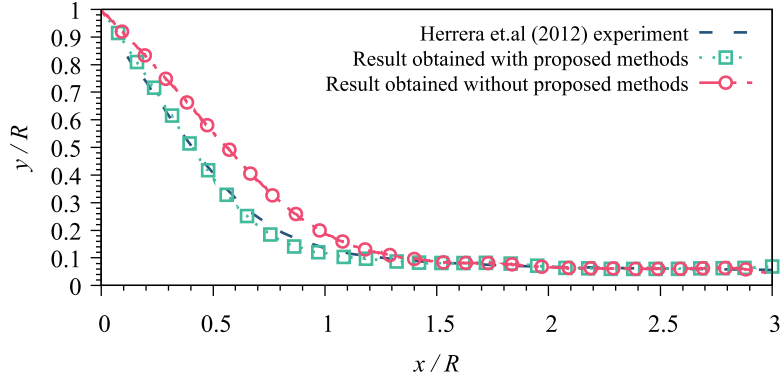}
\caption{\label{fig::jetOutline} The stable cone-jet shape obtained in the present study compared with experimental results in literature. }
\end{figure}

Fig.~\ref{fig::jetProcess} depicts the jet formation process and gives the corresponding velocity distribution. The strong electric field around the tube deforms the liquid drop into a cone-like shape, and the jet is ejected from the tip of the cone. With the increase of jet length, the end of the jet becomes unstable and breaks down to form droplets. The vortex flow near the interface induced by the interface charge can also be observed in the cone. A steady state of cone-jet is reached when the flow rate from capillary tube equals the one released through the jet. Fig.~\ref{fig::jetOutline} illustrates the stable cone shape obtained by the present numerical simulations together with the experimental result of Herrera et.al\cite{Herrera1}. The outline of the cone-jet obtained by the proposed face discernment and flux correction method agrees well with the experiment result while the simulation without the proposed methods provides a cone shape deviated from the experimental data.

\begin{figure}[htbp]
	\centering
	\subfigure{
		\label{fig::cone_rhoe_limit}
		\includegraphics[scale=0.4]{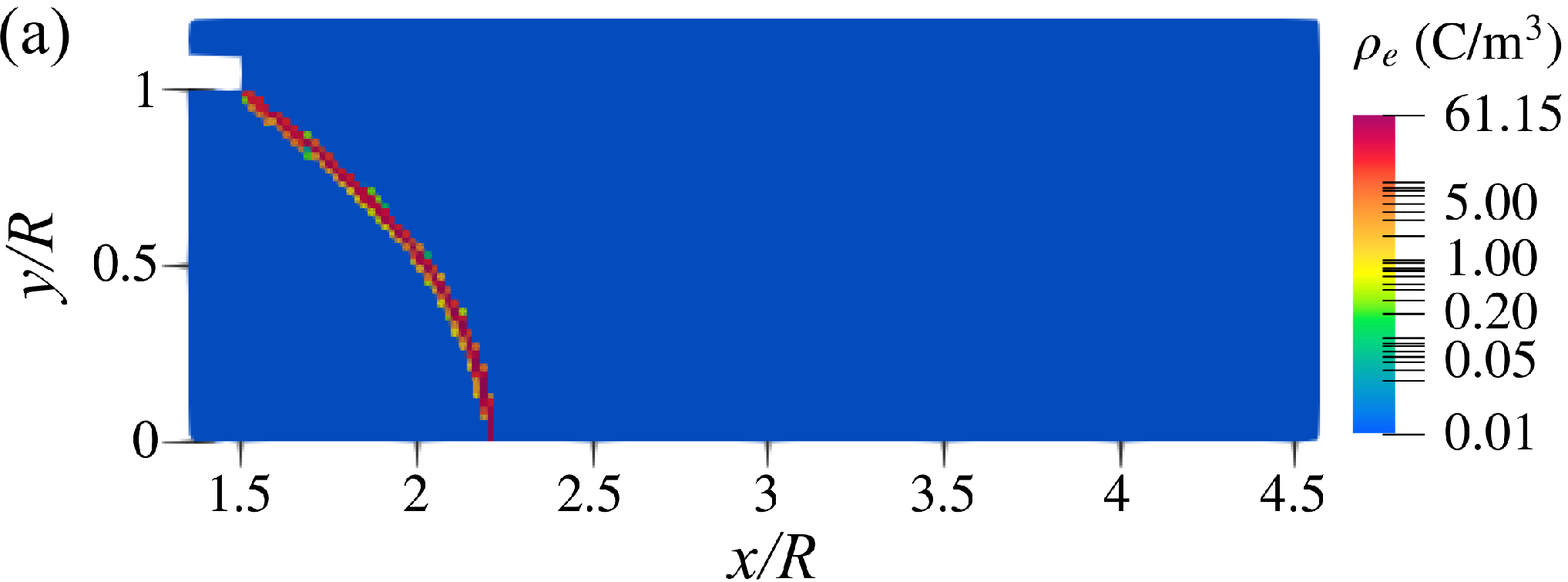}
	}
	\subfigure{
		\label{fig::cone_u_limit}
		\includegraphics[scale=0.4]{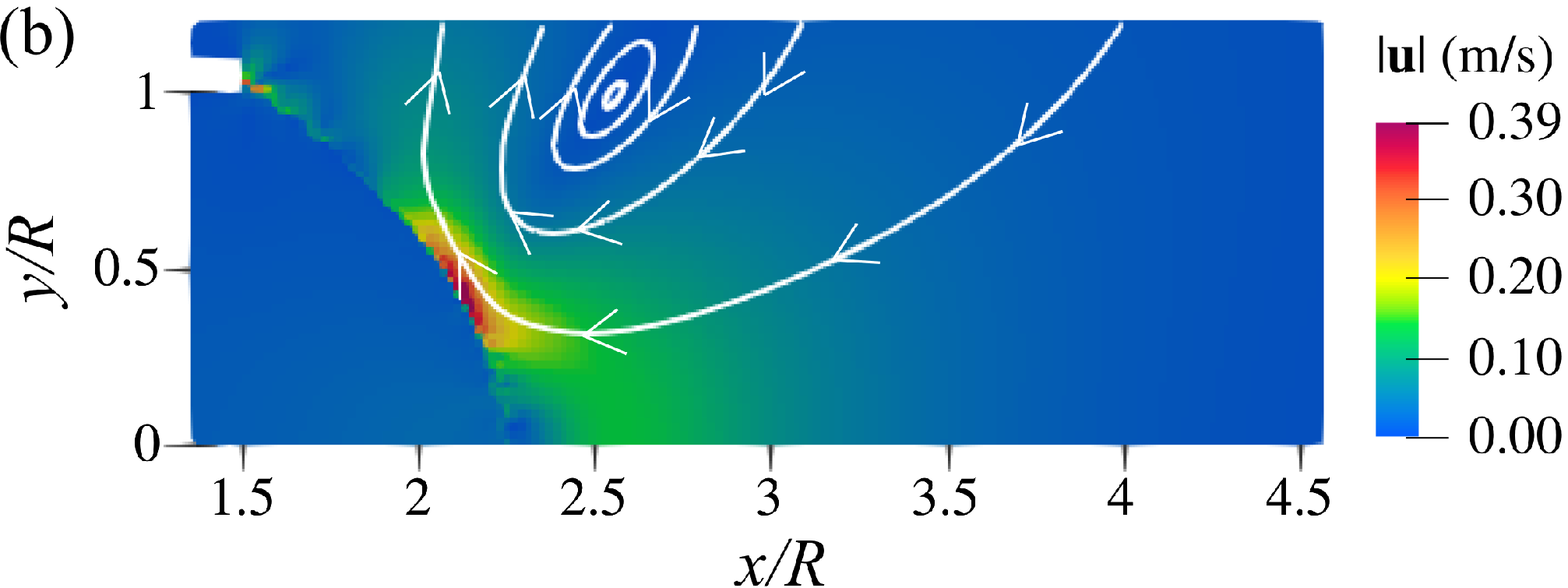}
	}
	\subfigure{
		\label{fig::cone_rhoe_nolimit}
		\includegraphics[scale=0.4]{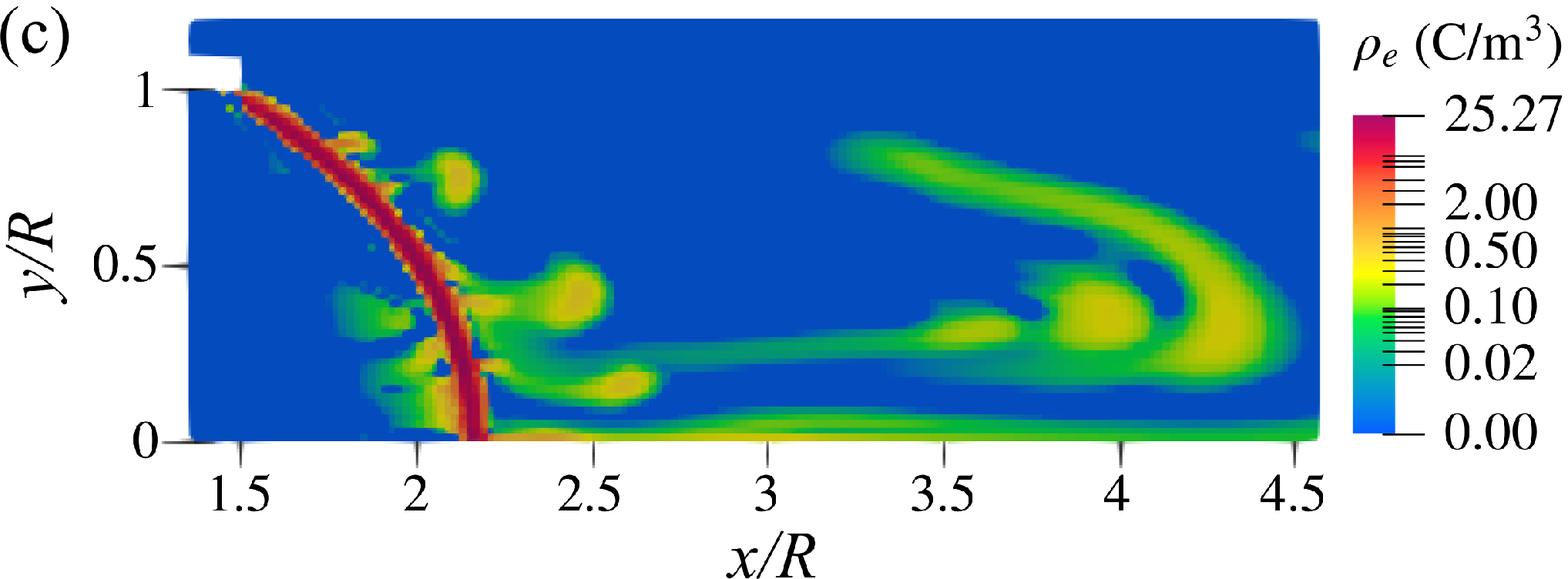}
	}
	\subfigure{
		\label{fig::cone_u_nolimit}
		\includegraphics[scale=0.4]{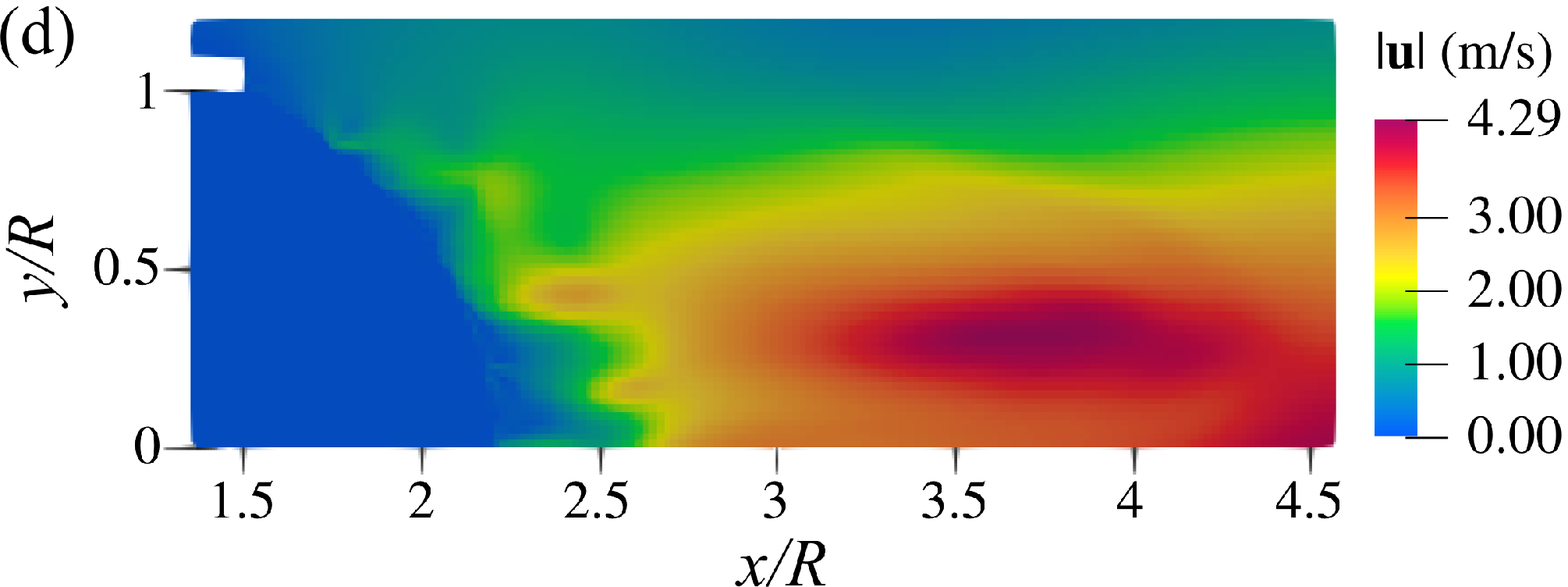}
	}
	\caption{\label{fig::uRhoeDistribution} The charge density and velocity distribution in the Taylor cone when $t=4\mathrm{ms}$. (a)(c) Charge density distribution. (b)(d)Velocity distribution. (a)(b) Results obtained with face discernment and flux correction method. (c)(d)Results obtained without face discernment and flux correction method.
	}
	
\end{figure}

The difference between the experimental cone shape and the simulated cone shape without face discernment and flux correction method is due to the charge leakage near the Taylor cone. Fig.\ref{fig::uRhoeDistribution} depicts the velocity and charge density distribution with and without the proposed methods during the simulation. The result without proposed method shows an obvious charge leakage in air, which finally decreases the charge density as well as the Coulomb force near the interface. The weakened Coulomb force makes the stress balance at the interface deviate from the real situation, resulting in the difference between the numerical and the experimental results. The leaked charge also causes a huge unphysical velocity in air phase due to the effect of Coulomb force. While in the result with proposed schemes, the velocity in the air is mainly induced by the tangential electric force at the interface, which is similar to the vortex flow in Fig.\ref{fig::deformationVelocity} and Fig.\ref{fig::jetProcess}.

The maximum velocity in the case where the face discernment and flux correction method are absent expands several times copmared with the no leakage result. The huge abnormal nonphysical velocity will seriously affect the stability of the simulation and thus a limitation to the Courant number $Co$ and interface Courant number $Co_\alpha$ is necessary. The $Co$ and $Co_\alpha$ are defined as:
\begin{equation}
C o=\frac{\Delta t \sum_{f}\left|\mathbf{u}_{\mathbf{f}} \cdot \mathbf{S}_{\mathbf{f}}\right|}{2 V_{p}}
\
,
\label{eq::Co}
\end{equation}
\begin{equation}
C o_{\alpha}=\omega C o
\
.
\label{eq::alphaCo}
\end{equation}
Here, $\omega$ is a scalar field whose value equals to 1 in the region where $1<\alpha<0.99$. In practice, the total time cost for a cone-jet simulation(0ms-10ms) using 4 cores in an intel$^{\circledR}$ Xeon$^{\circledR}$ E5-2650 V4 processor are 33820.2s and 68972.2s for the case with and without proposed methods , respectively, which shows a strong advantages for the proposed face discernment and flux correction method in computing resource consumption.

\section{\label{sec:conclusion}Concluding remarks}
In multiphase EHD problems, the charges in the conducting phase might be leaked into insulating phase during the simulation due to the numerical asynchronous transportation between the interface and charge, which will wreck the accuracy of the simulation results and increase the simulation cost. To avoid this unphysical error, two innovative schemes named face discernment method and flux correction method are proposed in this study to guarantee the single-phase transportation feature during the numerical simulation.

The aim of the face discernment method is to guarantee an accurate charge transportation through ohmic conduction by providing an accurate distribution of the physical properties. To achieve this goal, all of the possible ways for the interface to pass through the mesh cell faces are considered in the face discernment method and the phase fraction as well as the physical properties near the interface are reset according to the interface position in a cell. The flux correction method is designed to modify the velocity flux to prevent the charge from crossing the interface through convection. A single-phase correction step is firstly applied in this method to migrate the charge in the conducting phase, and then a addition correction flux with an adaptive correction factor is introduced to bring back the leaked charged back into the conducting media. These two methods only require a volumetric phase distribution and thus are easy to implement to other platforms. The whole algorithms are available to download and released as open source.

Several cases are introduced to test the overall performance of the proposed algorithms. The accuracy, single-phase constriction feature as well as the conservation of the methods are carefully validated, and satisfying performance of the face discernment and flux correction method is always obtained in the test. A Taylor cone-jet is finally presented to check the algorithms in real application scenarios. The results show that leaked charge results in an imprecise stress balance at the interface, leading to a different cone shape compared with the experimental results. The proposed methods can successfully prevent the charge leakage and eliminate the unphysical velocity in the air, which will improve both the accuracy and efficiency of the simulation.

Similar single-phase transportation requirement is common in the chemical engineering and other research community as shown in Sec.\ref{sec:introduction}. Since the charge transportation equation can be regarded as a scalar transport equation with the ohmic conduction term working as an explicit source term or diffusion term, the proposed methods should also be applicable to these problems which involve scalar transportation in the multiphase system. Using the face discernment and flux correction method to solve the challenging problems in these areas can be a future work.

\section*{Acknowledgments}
Jian Wu acknowledges financial support by the National Natural Science Foundation of China (Grant No.12172110), by the National Key R\&D Program of China (Grant No.2020YFC2201000) and the Fundamental Research Funds for the Central Universities (Grant No.AUGA9803500921). Jie Zhang acknowledges financial support by the National Natural Science Foundation of China (Grant No.11872296)

\section*{References}


\end{document}